\documentclass[acmlarge]{acmart}
\AtBeginDocument{%
  \providecommand\BibTeX{{%
    \normalfont B\kern-0.5em{\scshape i\kern-0.25em b}\kern-0.8em\TeX}}}

\setcopyright{acmcopyright}
\acmJournal{IMWUT}
\acmYear{2022} \acmVolume{6} \acmNumber{2} \acmArticle{64} \acmMonth{6} \acmPrice{}\acmDOI{10.1145/3534590}

\usepackage{caption}
\usepackage{subcaption}
\usepackage{tabularx}
\usepackage{algorithm}
\usepackage{algpseudocode}

\usepackage[utf8]{inputenc}

\begin{document}

\title{Modeling the Noticeability of User-Avatar Movement Inconsistency for Sense of Body Ownership Intervention}

\author{Zhipeng Li}
\email{lzp20@mails.tsinghua.edu.cn}
\orcid{0000-0001-6602-0176}
% \affiliation{%
%   \city{Beijing}
%   \country{China}
%   \postcode{100084}
% }
\author{Yu Jiang}
\authornote{This work was done while Yu Jiang was an intern at Tsinghua University.}
\orcid{0000-0001-7111-2374}
% \affiliation{%
%   \city{Beijing}
%   \country{China}
%   \postcode{100084}
% }
\author{Yihao Zhu}
\orcid{0000-0002-6620-3420}
% \affiliation{%
%   \city{Beijing}
%   \country{China}
%   \postcode{100084}
% }
\author{Ruijia Chen}
\orcid{0000-0002-1655-6228}
% \affiliation{%
%   \city{Beijing}
%   \country{China}
%   \postcode{100084}
% }
\author{Ruolin Wang}
\orcid{0000-0003-1445-716X}
% \affiliation{%
%   \city{Beijing}
%   \country{China}
%   \postcode{100084}
% }

\author{Yuntao Wang}
\orcid{0000-0002-4249-8893}
% \affiliation{%
%   \city{Beijing}
%   \country{China}
%   \postcode{100084}
% }

\author{Yukang Yan}
\orcid{0000-0001-7515-3755}
\authornote{Denotes as the corresponding author.}
% \affiliation{%
%   \city{Beijing}
%   \country{China}
%   \postcode{100084}
% }

\author{Yuanchun Shi}
\orcid{0000-0003-2273-6927}
\affiliation{%
  \institution{Tsinghua University}
  \city{Beijing}
  \country{China}
  \postcode{100084}
}

\renewcommand{\shortauthors}{Li et al.}

\begin{abstract}
An avatar mirroring the user’s movement is commonly adopted in Virtual Reality(VR).
Maintaining the user-avatar movement consistency provides the user a sense of body ownership and thus an immersive experience. 
However, breaking this consistency can enable new interaction functionalities, such as pseudo haptic feedback~\cite{majed2019pseudo} or input augmentation~\cite{wentzel2020improving, ivan1996go}, at the expense of immersion.
We propose to quantify the probability of users noticing the movement inconsistency while the inconsistency amplitude is being enlarged, which aims to guide the intervention of the users' sense of body ownership in VR.
We applied angular offsets to the avatar's shoulder and elbow joints and recorded whether the user identified the inconsistency through a series of three user studies and built a statistical model based on the results.
Results show that the noticeability of movement inconsistency increases roughly quadratically with the enlargement of offsets and the offsets at two joints negatively affect the probability distributions of each other.
Leveraging the model, we implemented a technique that amplifies the user's arm movements with unnoticeable offsets and then evaluated implementations with different parameters(offset strength, offset distribution).
Results show that the technique with medium-level and balanced-distributed offsets achieves the best overall performance.
Finally, we demonstrated our model's extendability in interventions in the sense of body ownership with three VR applications including stroke rehabilitation, action game and widget arrangement.
\end{abstract}




\begin{teaserfigure}
  \centering
  \includegraphics[width=0.8\textwidth]{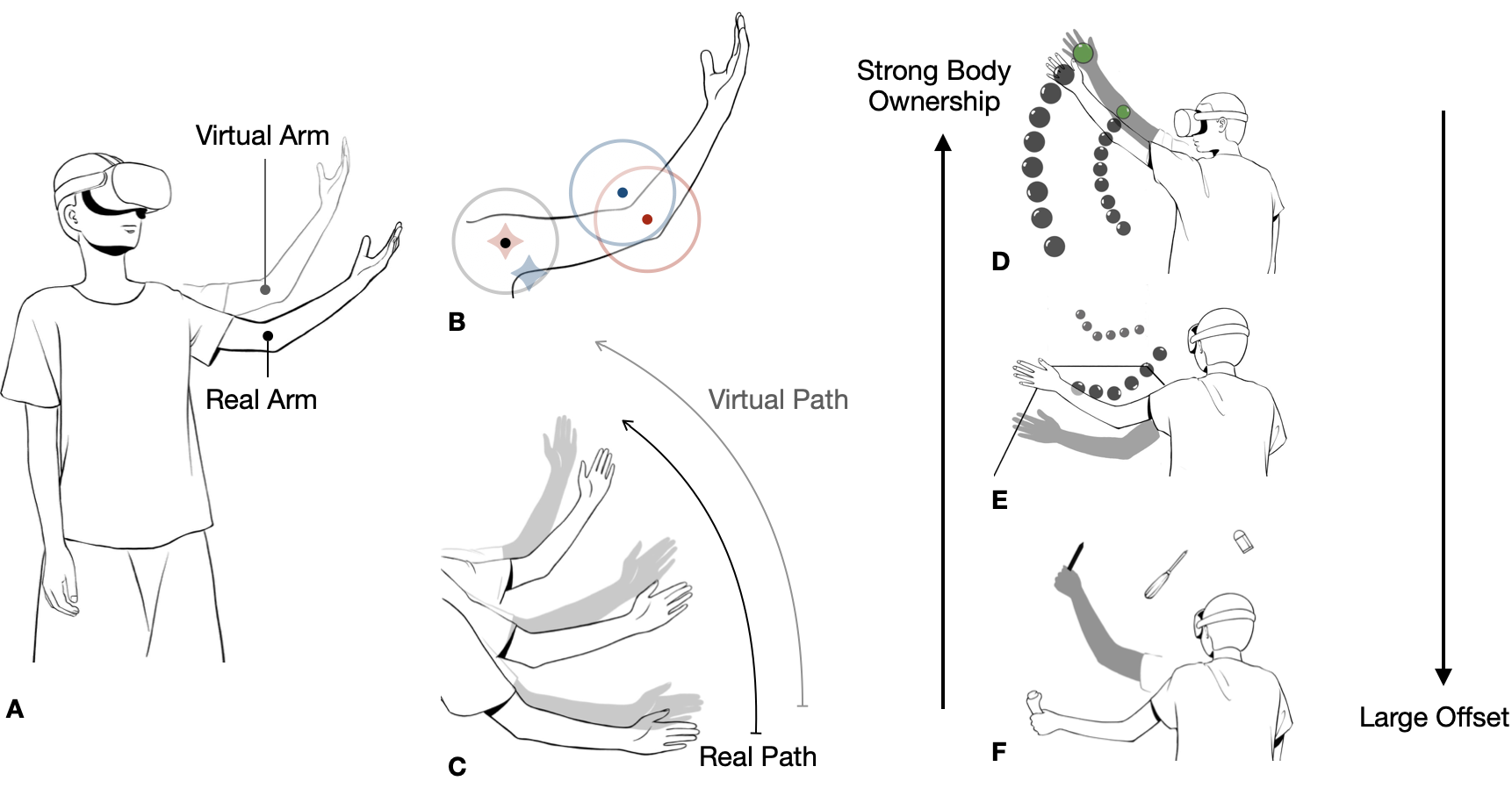}
  \caption{A: We investigate the effect of user-avatar movement inconsistency on body ownership; B: We apply angular offsets to the shoulder and the elbow joints and measure the probability of the user noticing the offset; C: We develop a model that dynamically calculates a set of applicable offsets given a requirement for noticing probability; D: An rehabilitation application of creating illusion of motor performance improvement with high body ownership E: A game application of changing physical effort demand by applying offsets at some expense of body ownership, and F: A target selection application of sacrificing body ownership for augmenting the user's input and thus reducing the physical burden.}
  \Description{This is a teaser figure and based on labeled drawings. The left shows a person wearing a head mounted display with his arm holding out and labeled “real arm”. A grey arm overlays at his arm but at a higher position is labeled “virtual arm”. The top figure in the middle zooms in to the left arm with a grey hollow circle centered at his shoulder with the center labeled with a black dot. A red star is at the center and a blue star is at its down right direction touching the edge of the circle. A red dot labels the elbow’s center as well as the center of a hollow red circle sized the same with the grey circle. A blue hollow circle and its centering blue dot is at the upper left direction of the red dot. The distance between the blue and the red dots is the same the the distance between the blue and the red stars at the shoulder. The bottom figure in the middle illustrates three arms, one higher than another and three grey arms each higher than one the black outlined arm. The higher black-grey arms has larger distance between them. In the right figure, the axis at the left pointing upwards writes “Strong Body Ownership” at the top; the axis at the right point downwards writes “Large Offset” at the bottom.}
  \label{fig:teaser}
\end{teaserfigure}

\maketitle

\section{Introduction}
\label{section:introduction}

While wearing a Virtual Reality (VR) head-mounted display (HMD), the user cannot see their own body in the physical world.
Instead, they usually embody virtual avatars mirroring their motions from a first-person perspective.
In this manner, they have the sensation that their own body is substituted by the avatar and receive a strong sense of body ownership ~\cite{Slater2010, Kilteni2012, dongsik2017theimpact}.
The motion consistency between the user and the avatar is essential to maintaining immersion in VR. 
However, breaking the consistency can enable new interactions with powerful functionalities, including input augmentation~\cite{roberto2017ergo, mcintosh2020iteratively} and pseudo-haptic feedback such as weight simulation in VR~\cite{majed2019pseudo} at the expense of body ownership.
Yet, there lacks a thorough understanding of how this cost of body ownership changes as the movement inconsistencies become more significant, or more fundamentally, how the noticeability of the inconsistency reacts to the amplitude enlargement.
This knowledge can guide interaction technique implementations to maintain the sense of body ownership to the extent required for an immersive experience while exploiting movement inconsistency for novel interactions.

We explored arm movement inconsistency as it is one most agile and frequently-used body parts (Fig~\ref{fig:teaser}A).
Since movement can be dissected into a series of poses, we start by investigating pose inconsistency, which can be implemented by applying angular offsets to the arm joints.
Instead of directly modeling the effects of inconsistency on the sense of body ownership, we decide to first investigate on our first research question \textbf{RQ1}: how strength of arm movement inconsistency affect the probability of it being noticed.
We then hope to leverage this knowledge to provide guidelines on the applicable range of movement inconsistency that be highly useful in supporting applications with various levels of the sense of body ownership requirements.
This, therefore, raises our second research question \textbf{RQ2}: how to leverage applicable inconsistency in building interaction techniques with different body ownership requirements?

To answer \textbf{RQ1}, we conducted three user studies to quantify the noticeability of movement inconsistency on single and both arm joints in various directions.
We first investigated varying the strength of offset on single joint at orthogonal axes separately. 
Then we explored how offset direction influences the noticeability. 
Finally we quantify the offset noticeability for composite two-joint offsets (Fig~\ref{fig:teaser}B). 
The user studies show that inconsistency noticeability increases roughly quadratically with stronger offsets yet the effect vary across axes, with a $13.31$ degree offset strength having an average $50\%$ inconsistency noticing probability, and it remains roughly the same at different offset directions. 
Finally, the probability distributions at two joints interacts negatively with each other.

In answering \textbf{RQ2}, we leveraged the results to construct a statistical model which outputs a set of applicable composite two-joint offsets given the requirement for noticeability and a arm pose. 
Based on the model, we proposed an example optimization goal of continuous amplification and implemented an interaction technique leveraging offset-applied movement with a maximum $75\%$ offset noticeability (Fig~\ref{fig:teaser}C). 
To be noted, other than this example, the model has the potential to support building various applications with different body ownership requirement and different optimization goals.
We evaluated the technique and compared how offset level and offset distribution between two joints affect the efficiency of task completion and interaction experience. 
The results show that the offset could significantly reduce the moving time and distance while a medium-level offset does not compromise the sense of body ownership.
Offset distribution between the shoulder and the elbow doesn't affect interaction efficiency, yet a balanced distribution improves interaction experience and helps maintain body ownership.

To demonstrate the extendability of our model in intervention in the sense of body ownership, we developed three applications that have different requirements for body ownership. 
The motivating stroke rehabilitation application aims to provide a body illusion that requires high body ownership. We thus apply unnoticeable offsets to the virtual avatar's movement so that the user perceives a motor performance improvement virtually and therefore gains stronger engagement~\cite{hee2020rehab} (Fig~\ref{fig:teaser}D);
For the VR action game, which requires less body ownership, we adjust the level of the offset to change the demand for physical effort from users(Fig~\ref{fig:teaser}E); 
Augmenting input on tasks such as widget arrangement sacrifices body ownership for augmentation functionalities. We thus apply strong offsets to amplify the user's motion to reduce users' physical burden (Fig~\ref{fig:teaser}F). 

This paper's contributions are three-fold:
\begin{itemize}
    \item We thoroughly investigated the statistical relationship between the applied offset and its noticeability on static poses. 
    Results show that the noticeability rises as the offsets increase in strength with roughly the same tendency for offsets in different directions. 
    The offsets on the shoulder and elbow interact negatively with each other.
    \item We leveraged the obtained knowledge to build a statistical model which predicts a set of applicable offsets on the shoulder and the elbow when given a noticeability requirement and a arm pose as inputs. We will open-source our model upon acceptance of this work.
    \item We implemented an input augmentation technique with the model. 
    We evaluated the technique and developed three applications with different body ownership requirements to demonstrate the extendability of the model.
\end{itemize}

\section{Related Work}
\label{section:related}

\subsection{Body Ownership in VR}
The illusion of body ownership refers to a person considering a non-bodily object as part of their own body. 
One of the most famous studies is the rubber hand illusion~\cite{Guterstam2011}, where the user has an illusion that the rubber hand is part of their body and reacts strongly to any harm done to the rubber hand.

VR can provide an immersive experience of body ownership illusion~\cite{Slater2010, slater2014first, mel2009inducing} as well as affecting the user's cognition~\cite{Peck2013}, perception~\cite{Van2011}, emotion~\cite{Jun2018} and behavior \cite{Banakou2014}. 
For instance, Jun et al.~\cite{Jun2018} investigated whether the virtual avatar's facial expression can modulate the user's emotion with the illusory feeling of full-body ownership of a virtual avatar.
Leveraging motion capture systems~\cite{Optitrack, Xsens} researchers reconstruct the user's physical motion and render the virtual avatar mirroring the motion to provide a strong sense of body ownership~\cite{slater2006a, pan2011confronting, Yee2007, Kilteni2012, Schwind2017}.
Since the arm is one of the most flexible and functional body parts, a number of works tried to provide the illusion of body ownership on a virtual arm~\cite{lorraine2016need, ryan2019virtual, ke2015the, ma2013the}. 
Lorraine et al.~\cite{lorraine2016need} investigated how the appearance of a virtual hand affects the own-body perception and found that a realistic human hand model elicits the strongest illusion. 
Ryan et al.~\cite{ryan2019virtual} investigated the influence of different hand visualizations on body ownership of grasping objects in VR, and the optimal visualization was to not penetrate the object while grasping, which complies with reality.

Most of the previous research focused on the precision of reconstructing and visualizing the user's real motion to elicit body ownership illusion. 
However, research on what precision level is sufficient for providing the illusion of body ownership and how differences between physical and virtual motions affect the illusion is very limited.
In this paper, we intentionally add motion differences and quantify their influences on the user's illusion of body ownership.

\vspace{-1mm}
\subsection{Modifying the Virtual Motion}
Instead of reconstructing the motion as precisely as possible, previous research also intentionally broke the consistency for useful functionalities in VR, including input augmentation, redirect walking, and pseudo-haptic feedback.

One of the most widespread applications is input augmentation. Researchers leveraged the modified movement to change the movement speed or the arm length~\cite{mcintosh2020iteratively, ivan1996go, wentzel2020improving, li2018evaluation, scott2007prism}. 
The Go-Go~\cite{ivan1996go} technique extends the position of the hand in-depth non-linearly, enabling users to interact with objects beyond reach. 
The PRISM technique~\cite{scott2007prism} applied an offset dynamically based on the user's hand speed to reach distant objects.
However, these techniques prioritize input augmentation and neglect the reduction in body ownership. 
Li et al.~\cite{li2018evaluation} compared four different motion amplification methods and found that large offsets reduced the feelings of immersion and body ownership.
Ownershift~\cite{tiare2018ownershift} proposed to slowly put the user's physical hand down while the virtual hand remains high for interaction. 
Thus the user could maintain ownership of the virtual hand while with the reduced arm fatigue.
Wentzel et al.~\cite{wentzel2020improving} proposed to amplify the hand position in-depth with an adaptive function that keeps the user's reach within reasonable bounds. Thus it could increase physical comfort while maintaining task performance and body ownership. 

On the other hand, a line of research on redirect walking~\cite{rietzler2020telewalk, sharif2001redirected, frank2010estimation} modifies the perceived forward direction of users so that they feel like walking straight while actually advancing in circles. 
This expands a finite space in reality with infinite virtual space. 
Rietzler et al.~\cite{rietzler2020telewalk} proposed to modify the user's head rotation to enable directional changes without any physical turns. Thus the user could always be on an optimal circular path inside the real world while walking freely inside the virtual world. 

Another functionality is the providence of the pseudo-haptic feedback~\cite{anatole2009simulating, anatole2000pseudo, mahdi2016haptic, andreas2009hemp, andreas2008hemp}.
The offset between the real and the virtual hand produces the illusion of stiffness while the user is pushing against a virtual object~\cite{anatole2000pseudo}. 
Pusch et al.~\cite{andreas2009hemp, andreas2008hemp} used visual hand displacements to create a feeling of wind resistance.
It can also be used to simulate the weight of objects in VR~\cite{david2014toward, nakakoji2010Toward, palmerius2014An, majed2019pseudo, dominjon2005influence, michael2018breaking}. For instance, Samad et al.~\cite{majed2019pseudo} leveraged the ease in moving light objects compared to heavy ones and manipulated the control-display ratio between the user's movement and the visualization of the movement to induce an illusory perception of weight. 

These techniques show that breaking the movement consistency can enable new interaction functionalities at the expense of body ownership.
However, researchers apply offsets with their own purposes but have limited understanding of the cost of body ownership, to which extent the users notice the inconsistency. 
Thus we propose to quantify the effect of movement inconsistency on the noticeability and finally the sense of body ownership. 
We expect the proposed model to serve as a tool for designing and implementation interaction techniques with different requirements.

\section{Methodology}
\label{section: methodology}

We term user-avatar movement inconsistency as the differences between the key joint positions of the user's physical body and the avatar's virtual body.
To understand how the inconsistency influences the user's sense of body ownership, we intentionally add offsets to the avatar body and measure how the possibility of the user noticing the difference increases as we enlarge the offset.
Specifically, we define an offset as noticeable when users can detect the movement inconsistency and do not perceive the avatars' bodies as their own when it is added. 
In this section, we introduce the methodology to quantify the relationship between the movement inconsistency and the sense of body ownership step by step.
Before that, we first formulate the problem and discuss the hypothesis.

\subsection{Selecting the Left Shoulder and Elbow for Investigation}
Based on existing research, we extend to study the motion modification on more than one key joint.
Before being able to conduct full-body motion modification, which requires controlling a very large number of degrees of freedom, we decided to investigate adding offsets on limbs composed of several connected key joints as the first step.
We chose the shoulder and the elbow of the left arm to apply offsets. Since the arm is one of the most agile and frequently-used body parts, we assume that the user will be more sensitive towards the movement inconsistency of the arm compared to other body parts. 
We expect the approach on these two key joints to be applicable and extendable to other body parts, potentially containing more than two joints.

\subsection{Defining an Arm Pose}
\label{section: parameterize}

\begin{figure}[htbp]
    \begin{minipage}{0.4\linewidth}
        \centerline{\includegraphics[width=\textwidth]{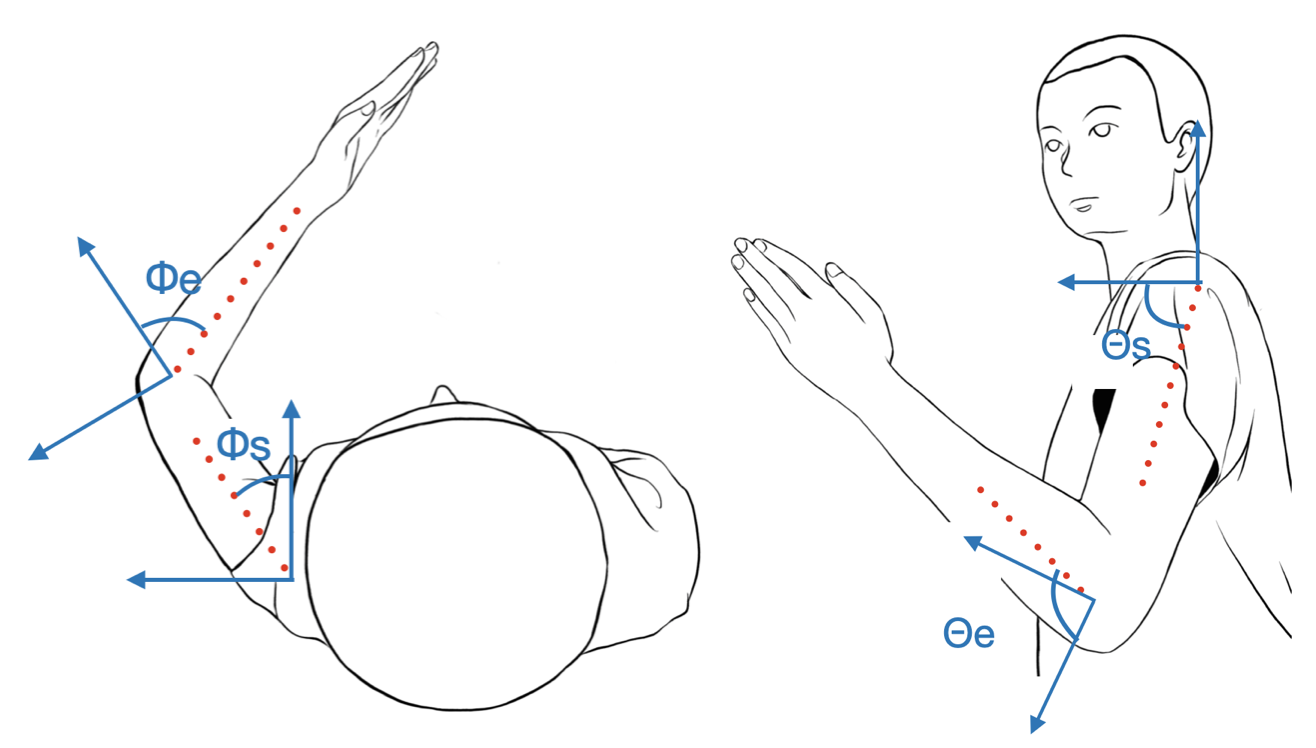}}
        \caption{The spherical polar coordinate systems where we calculate the shoulder and the elbow joint coordinates. The coordinate system of the shoulder is relative to the body and that of the elbow is relative to the upper-arm. }
        \label{fig: pose_para}
    \end{minipage}
    \hspace{1.5cm}
    \begin{minipage}{0.3\linewidth}
    	\centering
    	\captionof{table}{Symbols used to represent the axis angles and offsets.}
        \begin{tabularx}{\textwidth}{cccc}
            \toprule
            Position & Axis & Angle & Offset \\
            \midrule
            Shoulder & $\phi$ & $\Phi_{s}$ & $\phi_{s}$ \\
            Shoulder & $\theta$ & $\Theta_{s}$ & $\theta_{s}$ \\
            Elbow & $\phi$ & $\Phi_{e}$ & $\phi_{e}$ \\
            Elbow & $\theta$ & $\Theta_{e}$ & $\theta_{e}$ \\
            \bottomrule
            \raggedbottom
        \end{tabularx}
    \vspace{20pt}
    % \captionof{table}{Symbols used to represent the axis angles and offsets.}
    \label{tab: para_all}
    \end{minipage}
\end{figure}

As previous studies explored the effect of scaling the lengths of the upper arm and the forearm on embodiment~\cite{wentzel2020improving, mcintosh2020iteratively}, we focus on how rotational changes at the shoulder joint and the elbow joint affect the sense of body ownership. We thus employed the spherical polar coordinate system, which defines the position of a point with its distance to the origin $r$, its polar angle $\theta$, and its azimuthal angle $\phi$, instead of the Cartesian coordinate system, which represents a point in space with a triplet $(x,~y,~z)$. The former releases us from considering the length of the limb. As shown in Figure~\ref{fig: pose_para}, we define an arm pose by a pair of angular coordinate $(\Phi_\text{s},~\Theta_\text{s})$ and $(\Phi_\text{e},~\Theta_\text{e})$. The elbow coordinate is relative to the shoulder joint and the two axes are orthogonal with the upper-arm. We summarize our parameterization of the upper limb and the according offset in Table~\ref{tab: para_all}.  

\subsection{Hypothesis}
\label{section: hypothesis}
In investigating the effect of applying angular offsets on the sense of body ownership, we raise three hypotheses about the relationship between the interested variables, which were later testified by user studies. The hypotheses are as follow:
\begin{itemize}
    \item \textbf{H1: The stronger the offsets applied, the more likely users will notice the user-avatar inconsistency.} Smaller angular offsets might be unnoticeable, yet larger ones might result in significant and obvious changes in arm pose (e.g. curving an unbend real arm). We hypothesize that users are more likely to notice offsets when larger offsets are applied. 
    \item \textbf{H2: Users' sensitivity towards the offsets vary at different directions.} Offsets at different directions could affect how the user observes the arm pose and thus the user's sensitivity towards offsets.
    \item \textbf{H3: Users' sensitivities towards the shoulder and elbow joint offsets are dependent.} Since applying offsets at the shoulder passively affects the elbow position, we assume that shoulder offsets affect the sensitivity towards offsets at the elbow.
\end{itemize}

\begin{figure}[htbp]
    \centering
    \includegraphics[width=0.7\linewidth]{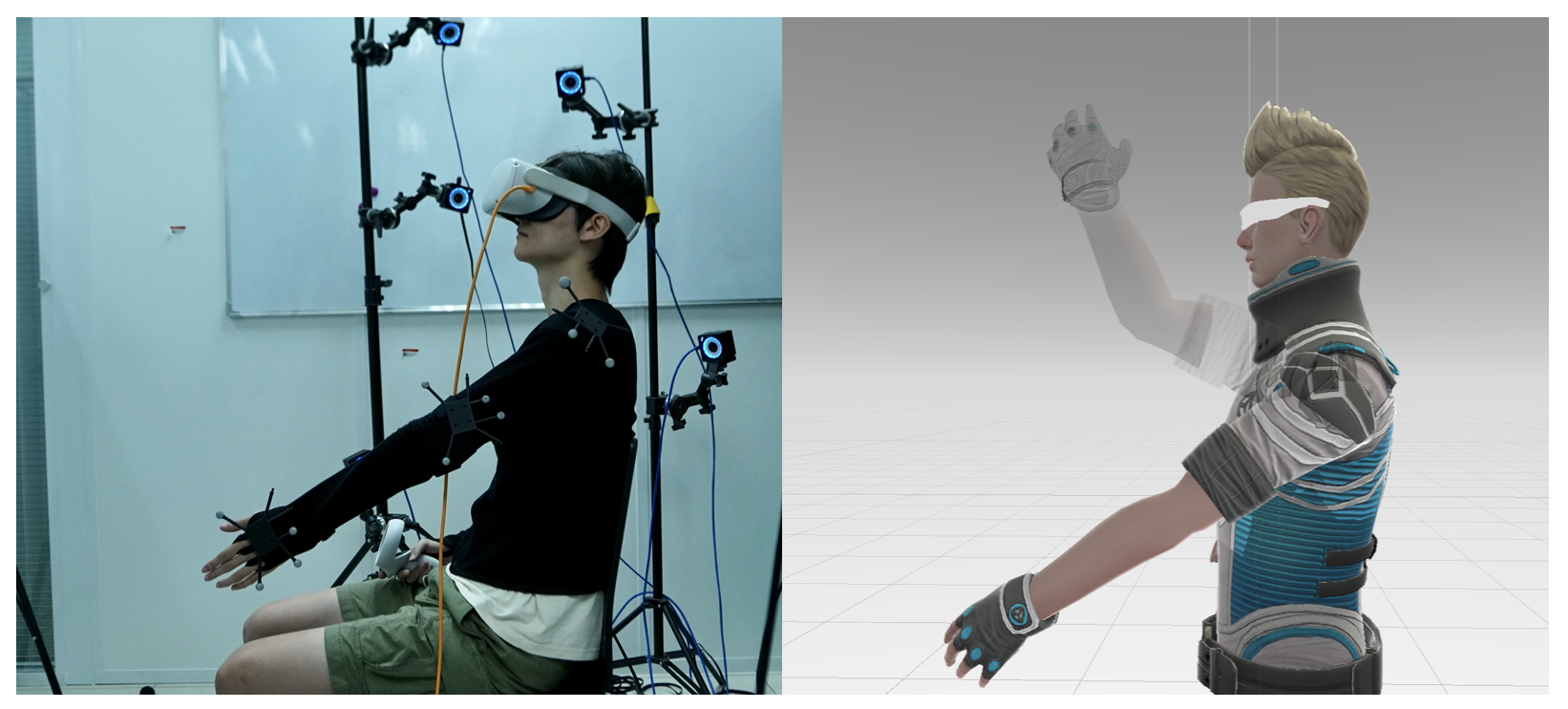}
    \caption{Left: the participant wears a bodysuit with three markers tracked by the Optitrack motion capture system. He wears the VR headset and sits on a comfortable chair during the study; Right: the experimental avatar from a third-person perspective. The virtual avatar mirrors the arm movement of the participant, and a semi-transparent arm is rendered to indicate the target arm pose.}
    \label{fig:procedure} 
\end{figure}

\vspace{-7mm}
\subsection{Task Design for Noticeability Test}
\label{section:experiment_procedure}

We designed a task to test whether the users lose their sense of body ownership when an offset is applied to the virtual arm of the avatar at a certain arm pose.
During each task, the position of the left shoulder and elbow of the user is tracked by a marker-based motion capture system. 
The virtual avatar's arm is constructed such that it deviates from the physical arm by a fixed angular offset no matter where the user moves their arm to.
Then to guide the user to perform a target arm pose, we render a semi-transparent virtual arm attached to the avatar's body and ask the user to adjust the virtual arm until it overlaps with the semi-transparent arm exactly.
After that, we remove the semi-transparent arm and ask whether participants perceive the virtual arms exactly as their own arms, as shown in Fig~\ref{fig:procedure}.
The participants are told to help evaluate the tracking precision of the motion capture system to avoid them suspecting the existence of offsets. 
As the output, we counted the number of participants who identified an offset between their physical arm pose and that of the virtual avatar for each offset. 
With an offset, we statistically simulate its offset noticing probability by the percentage of the participants who notice the inconsistency and leverage such probability as the metric for the strength of the sense of body ownership over the virtual avatar.

As the tasks test different offsets, directly reapplying the offsets onto the virtual arm results in visual glitches, which might impair body ownership. 
To avoid this, we asked users to close their eyes for a short rest while switching the tasks.
The system will play a prompt tone when the switch was done and users open their eyes to judge if the virtual arms are their own arms. 
If they recognize the virtual arm poses same as their own arms, they press button A on the controller held in their right hands and button B otherwise.
In a warm-up session before the experiment, participants perform ten tasks testing all arm poses without offsets, and
we orally reminded them to close their eyes after each task to help them get familiar with the procedure.
One of the most frequently used evaluations of body inconsistency noticeability is questionnaires \cite{tiare2018ownershift, botvinick1998rubber, kalckert2014moving}. 
However, the questionnaire is not appropriate in our experiment since users are asked to evaluate the body ownership instantly for every combination of pose and offset.
Therefore, we adapted a quantitative method to evaluate the body inconsistency noticeability which is similar to \cite{wentzel2020improving}.

\vspace{-2mm}
\subsection{Investigating Composite Two-joint Offsets}
\label{section: approach}

Considering the extremely large space of adding angular offsets (of different strengths and in different directions) to two joints in combination, we decide to investigate how the offsets influence body ownership in a bottom-up flow in four steps.

\textbf{In the first step}, we only added angular offsets to each joint's two orthogonal directions separately. Without the complication of different directions, this enabled us to test the strength of the offset with a resolution of 3 degrees in 11 levels.
We performed the noticeability tests in each level condition and precisely measured how offset strength related to the offset noticing probability. 
Then \textbf{in the second step}, we extended our investigation to offset directions and tested the directions in the entire space with a resolution of 15 degrees in 24 levels.
Considering the time and effort requirements, we shortened the testing range of offset strength to 12 to 24 degrees with the same resolution of 3 degrees.
Until then, we achieved an overall understanding of how angular offsets on a single joint affect their noticeability. 
So \textbf{in the third step}, we looked into the condition of two offsets being applied on the shoulder and the elbow at the same time.
Our goal was to model the relationship between the noticeability of composite two-joint offsets and the offset noticing probability at each joint.
However, the position of the elbow, as well as the forearm pose, are passively impacted by offsets applied at the shoulder. 
We thus first sampled the offsets with the same strength in 8 directions to apply to the shoulder and then tested the offsets of 4 strength levels and in 8 directions to apply to the elbow.
We observed how the composite two-joint offset's noticeability differs from that of offset applied only at the elbow to model their relationship further.
Then \textbf{in the final step}, based on the data we collected in the sampled angular offset space, we applied bi-dimensional linear interpolation to build a statistical model, which can calculate the probability of noticing the offset with a given composite two-joint offset and a pose.

\begin{figure}[htbp]
    \centering
    \includegraphics[width=0.75\linewidth]{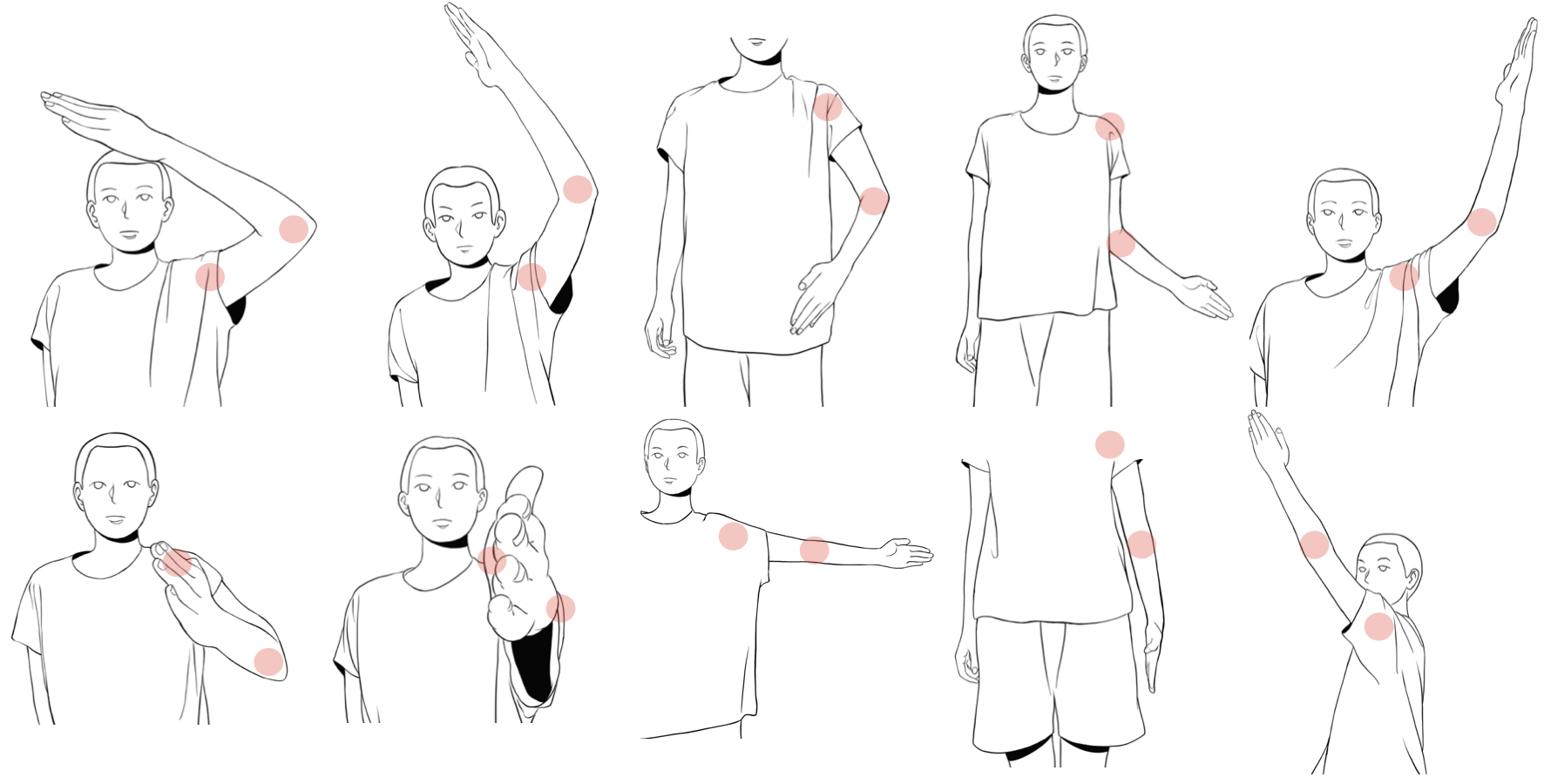}
    \caption{The 10 upper left arm poses tested in the following experiments, with the shoulder and elbow joints highlighted as red balls.}
    \label{fig:armposes}
\end{figure}

\subsection{Investigating the Effects of Arm Poses}
\label{section: clusterposes}

We consider the arm pose as an extra factor of the user's sensitivity to noticeability of inconsistency.
Through several pilot tests, we identified that compared to bent arm poses, users were more sensitive towards any offset applied to either joint when users raised their arms forward. 
So we also varied the arm poses, characterized by their joints' rotational angles, while adding each angular offset in the studies. 
Considering our limited capacity of human labor for data collection and the infeasibility of covering the entire space of human poses, we tested 10 different arm poses.
To ensure that the sampled poses are representative, we sampled 6 poses from CMU MoCap database~\cite{CMUMocap} consisting of motion-captured data of over $2500$ motion sequences and over $200000$ human poses with joint positions in real human activities. 
We added 4 poses to cover the extreme poses at which we believed the user would have the highest sensitivity.
We randomly sampled $50000$ frames and clustered them with the HDBSCAN algorithm~\cite{leland2017hdbscan} using the skeletal distance function given by Shakhnarovich et al.~\cite{shakhnarovich05learning}. 
Given $L$ joints in the pose, the distance between two poses $\alpha_1$ and $\alpha_2$ can be calculated by 

\begin{equation}
    \text{Distance}(\alpha_1, \alpha_2) = \max \limits_{1 \leq i \leq L} \sum_{d \in x, y, z} (\alpha^{i}_{d, 1} - \alpha^{i}_{d, 2})
\end{equation}

with $x, y, z$ being the locations of the joint. 
We added raising arm forward, raising arm upward, raising arm to the side, and putting arm downward as the extreme cases. 
Fig~\ref{fig:armposes} shows the $10$ sample poses with their parameters listed the supplementary material.

\section{Investigation of the two-joint offset noticing probability}
\label{section:study1}

In this section, we conducted three user studies to confirm whether our hypotheses are correct. Specifically, we investigated the effect of angular offset on offset noticing probability on two orthogonal axes at a joint, two-dimensionally at both joints, and on the two joints compositely.
All user studies had been approved by our university's IRB board.

\subsection{Phase 1: Investigating the Strength of the Offset}
\label{section:exp1}

We first investigated the effect of offset strength on the offset noticing probability by testing offset values of different strength on the four axes separately for the sampled $10$ poses. 

\subsubsection{Design}
\label{section: design}
The \textit{target poses} (Section \ref{section: clusterposes}) set as a control factor, which sample the space and are representative of human upper left limb poses. 
The independent variable is the \textit{single axis offset value}. 
The dependent variable is \textit{offset noticing probability} which we measure by the metric \textit{whether the offset is noticeable} with the applied offset on a pose. 
The study was conducted based on the procedure described in Section \ref{section:experiment_procedure}. 
We tested offset values ranging from $-15$ to $15$ degrees with a 3 degree interval for each of the four axes. 
Each participant thus evaluated $11 \textit{offset values} \times 4 \textit{axes} \times 10 \textit{target poses}= 440$ tasks in a random order.
Each participant's data collection was divided into 3 sessions with a five-minute break between sessions to avoid fatigue, lasting around 40 minutes in total. 

\subsubsection{Participants}
We recruited 12 participants from a local university(6 male, 6 female) with an average age of 21.15 (SD = 1.68). All participants were right-handed. The self-reported familiarity with VR score averages at 3.62 (SD = 1.26) with a 7-point Likert scale from 1 (not at all familiar) to 7 (very familiar).

\subsubsection{Apparatus}
\label{section: apparatus}

We developed a VR system to investigate whether an offset on a pose is noticeable, which was developed in Unity 2019 for the Oculus Quest2 headset powered by an Intel Core i7 CPU and an NVIDIA GeForce RTX 3080 GPU. 
The participant performed the experiment in a seated position and wore an adhesive bodysuit with three Optitrack rigid body trackers tagged at the left shoulder, elbow, and waist. 
We reconstructed the user movement based on the Optitrack data on a virtual humanoid (male or female) avatar with the user's viewpoint coinciding with the avatar's. 
An example scene is shown in Fig~\ref{fig:procedure}.

\subsubsection{Result}
\label{study1phase1result}

\begin{figure}
    \centering
    \includegraphics[width=0.95\linewidth]{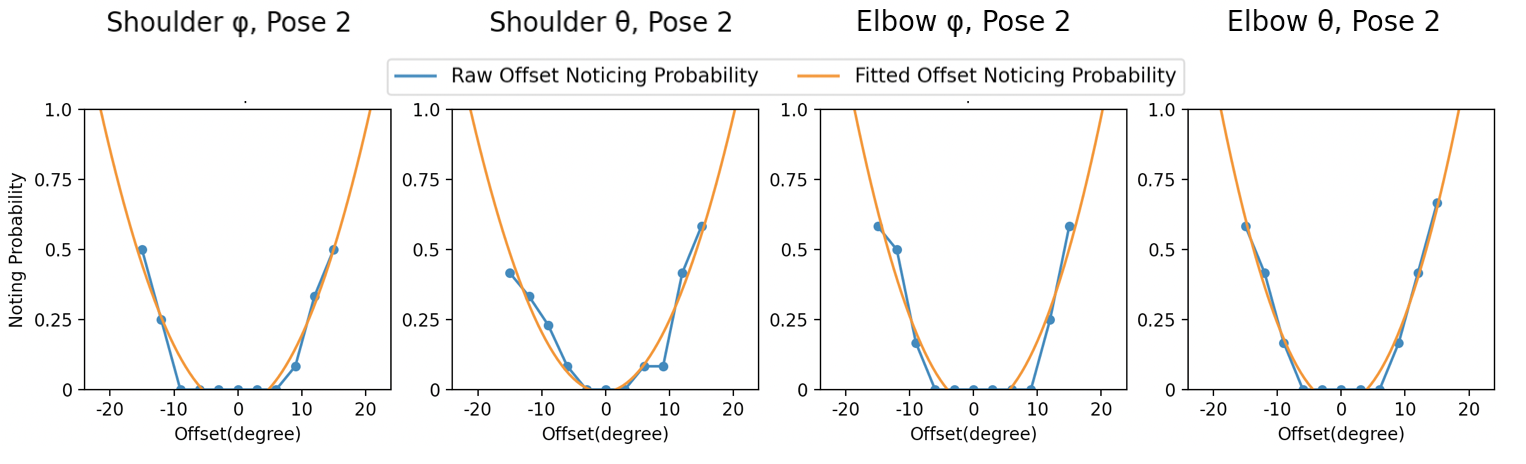}
    \caption{The offset noticing probabilities with offset values for the $\phi$ and $\theta$ axes at the shoulder and the elbow joints at Pose 2. The blue lines show the probabilities at the sampled offset values and the orange curves show the fitted quadratic curves.}
    \label{fig:exp1pose3}
\end{figure}

For each axis, we plotted the offset noticing probability with offset value for four axes for each pose. An example based on pose 2 is shown in Fig~\ref{fig:exp1pose3}. Please refer to the supplementary material for the complete result. For each subfigure, the x-axis is the offset value, and the y-axis is the probability of noticing the offset. We used a quadratic polynomial function to fit the data points with a root mean squared error of $0.034$. The figures demonstrate that \textbf{the offset noticing probability increases roughly quadratically with stronger offsets.} Also, the effect of offset strength on the offset noticing probability varies on each axis and for each pose. With the fitted relationship, the offset noticing probability given a single axis offset can thus be interpolated. The quadratic relationship confirmed the \textbf{H1}.

\subsection{Phase 2: Investigating the Direction of the Offset}
\label{section:exp2}
We then explore how the offset noticing probability is affected by the direction of the offset by testing two-dimensional offset values.

\subsubsection{Design}
The control factor and the dependent variable are kept the same with Section \ref{section: design} and the procedure similarly followed Section \ref{section:experiment_procedure}. 
We set the independent variable as \textit{two-dimensional offsets at a joint} which applies an offset with a certain value at a direction. 
We sampled two-dimensional offsets on circles with a radius ranging from $12$ to $24$ degrees with a $3$ degree interval and with a $15$ degree interval between neighboring points on each circle as shown in Fig \ref{fig:exp2circle}. 
The x-coordinate indicates the offset to be applied at the $\phi$ axis while the y-coordinate indicates the offset to be applied at the $\theta$ axis. 
The data range was chosen considering the large data size in sampling points two-dimensionally and our focus on investigating the offset noticing probability trend, which we could expect to be significant between $12$ to $24$ degrees as informed by the result in Phase 1 (Section~\ref{section:exp1}). 
Given the massive task number that one participant need to finish, we conducted tests at two joints for four poses to shorten the experiment time and reduce the influence of fatigue.
Base on the data collected on these four poses, we concluded the two-dimensional offset noticing probability distributions and validated the hypothesis.
Each participant thus performed $24 \text{ two-dimensional offsets per circle} \times 5 \text{ circles} \times 2 \text{ joints} \times 4 \text{ poses} = 960$ tasks in a random order. 
Each participant's data collection was divided into $8$ sessions with a five-minute break between sessions to avoid fatigue, lasting around $90$ minutes.

\subsubsection{Participants}
We recruited 12 participants including $6$ males and $6$ females. The participants' age ranged ranged from $20$ to $24$ $(AVG = 21.92,~SD = 1.57)$. All participants were right-handed. The self-reported familiarity score averages at $3.17$ (SD = $1.69$) with a 7-point Likert scale from $1$ (not at all familiar) to $7$ (very familiar).

\subsubsection{Apparatus}
The apparatus remains the same with Section \ref{section: apparatus}.

\subsubsection{Result}
\begin{figure}[htbp]
     \centering
     \begin{subfigure}{0.24\textwidth}
        \centering
        \vspace{25pt}
        \includegraphics[width=\linewidth]{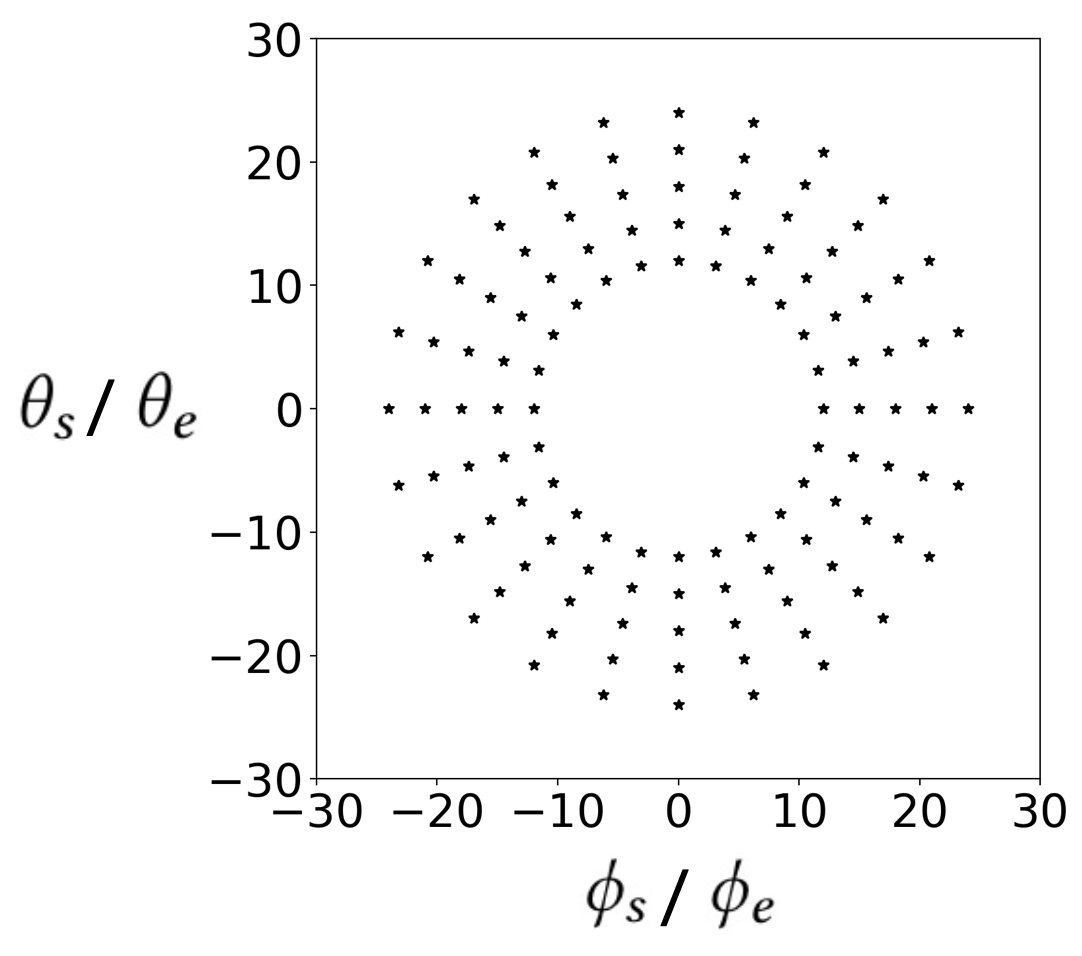}
        \vspace{7pt}
        \caption{ }
        \label{fig:exp2circle}
     \end{subfigure}
     \hfill
     \begin{subfigure}{0.65\textwidth}
        \centering
        \includegraphics[width=\linewidth]{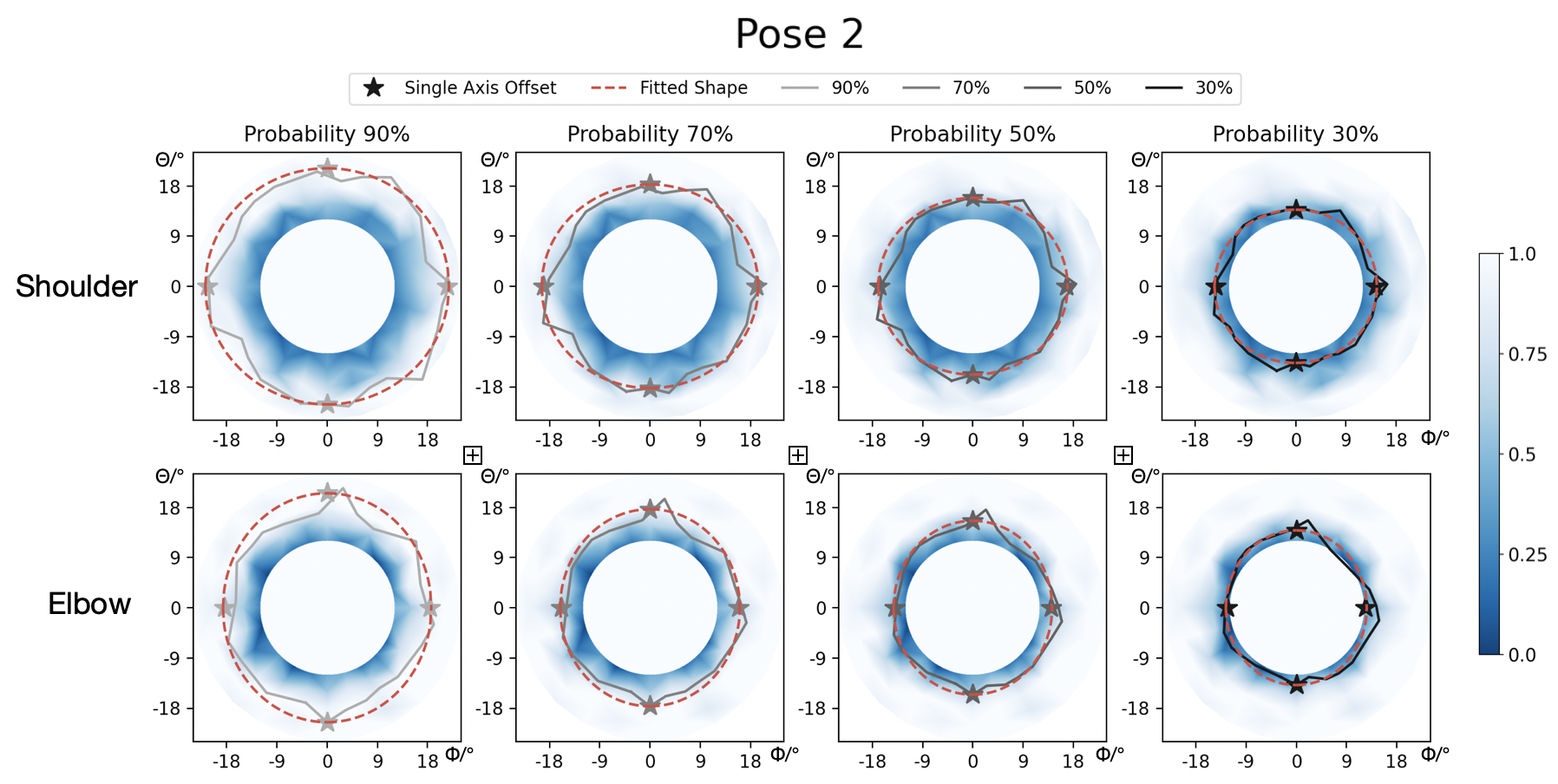}
        \caption{ }
        \label{fig:exp2ovals}
     \end{subfigure}
     \hfill
     \caption{(a) Sample points for the offset noticeable tests: we tested directions 15 degrees apart, and in each direction we tested points from 12 to 24 degrees with a 3 degree interval. (b)The relationship between two-dimensional offsets and the probability of noticing the offset at Pose 2. The heatmaps indicate the probability with darker color representing a lower noticing probability. Two-dimensional offset values with 90\%, 70\%, 50\%, 30\% noticing probabilities are shown. In each figure the black/grey oval indicates the raw offset value while the red oval indicates the fitted offset value at the probability. The stars in each figure illustrates the single axis offset values at that probability, interpolated from Section~\ref{section:study1}.}
\vspace{-2mm}
\end{figure}

We performed the linear interpolation on the collected data points to display heatmaps of offset noticing probability distributions. 
Fig~\ref{fig:exp2ovals} shows an example for the shoulder joint and elbow joint probability distributions for pose 2. 
Please refer to the supplementary material for the complete result. 
The x-axis and y-axis are the $\phi$ and $\theta$ offset values. 
The color intensity indicates the offset noticing probability with darker color indicating a lower probability in noticing the offset between visual and the real arm pose. 
Observe from Fig~\ref{fig:exp2ovals} that the shape of probability distribution on two-dimensional offsets roughly resembles an oval. 
Based on this data we can bi-dimensionally interpolate the offset noticing probability at a joint with a given two-dimensional joint offset. 
Similarly, the two-dimensional offset values at a given probability in each quadrant resemble a quarter oval bounded by the single-axis offset values at the same probability. 
This visual observation is uniform for two joints of all four poses we collected data on. 
With a given probability value, we thus fitted a quarter-oval curve at each quadrant with a function parameterized by its two single-axis offset values at each joint for each pose. 
Based on the fitted relationship, for pose $(p)$ and at joint $(j)$ the two-dimensional offset $(x,~y)$ at a probability of interest can be characterized by Equation~\ref{equ:ovalfunction} in different quadrants.
$\phi_{j \pm}$ and $\theta_{j \pm}$ represent the positive/negative single-axis offset values at that probability at the $\phi$ and $\theta$ axes respectively. 
For pose 2, the root mean square error was $2.03$ for the $90\%$ probability fitted curve, $1.34$ for the $70\%$ curve, $1.04$ for the $50\%$ curve, $1.03$ for the $30\%$ curve. 
The quadratic relationship rejected \textbf{H2} since \textbf{the offset noticing probabilities in different directions are roughly the same}.

\begin{equation}
    \text{Offsets on joint j of a given probability: } 
    \frac{x^2}{\phi_\text{j$\pm$}^2} + \frac{y^2}{\theta_\text{j$\pm$}^2} = 1
    \label{equ:ovalfunction}
\end{equation}

\vspace{-5mm}
\subsection{Phase 3: Investigating the Composite Two-joint Offset Noticing Probability}
\label{section:exp3}

Since applying offsets at the shoulder passively affects the elbow position, we thus hypothesized that the composite two-joint offset noting probability is dependent on such probabilities on the shoulder and elbow joints. In this section, we applied different offsets at the shoulder joint to observe how the probability at the elbow joint transforms.

\subsubsection{Pilot Study}
Due to the difficulty of large scale and dense data collection needed to investigate the two-joint offset noticing probability, we conducted a pilot study to gain more understanding into the transformation to aid targeted data collection in the formal study. We conducted a pilot study on $2$ poses with $4$ participants. We applied two shoulder joint offsets of different directions for each of the two poses. For pose 2, we applied offsets in the first and the third quadrants diagonally at 45 degrees. Respectively, for pose 2, we applied offsets in the second and the fourth quadrants. The value of the shoulder offset corresponds to $30\%$ offset noticing probability. We did noticeable offset tests for the elbow joint at points ranging from $9$ to $24$ degrees with a $3$ degree interval to explore the transformation. We recruited $4$ participants and each of them evaluated $2 \textit{shoulder offsets} \times 6 \textit{elbow offset sampling circles} \times 24 \textit{elbow offsets sampled per circle} \times 2 \textit{target poses}= 576$ tasks in a random order.

The pilot study's results are shown in Fig~\ref{fig:exp3pilot}. Similar to the previous study, the heat maps made using linear interpolation between data points represent the offset noticing probability. The yellow stars show that offset applied at the shoulder joint relative to the origin. 

Observe from the heat maps that the two-dimensional offset noticing probability distribution at the elbow joint shifts in the opposite direction of the shoulder joint offset direction. The shift amount is roughly the same as the shoulder joint offset amount. Meanwhile, the probability distribution pattern stays roughly the same.

\subsubsection{Design}
Based on the pilot study results, we conducted an experiment to further investigate the direction and amount of the elbow probability distribution shift.
The control factor and the dependent variable are kept the same with Section \ref{section: design} and the procedure similarly followed Section \ref{section:experiment_procedure}. 
We set the independent variable as a composite two-joint offset composed of two two-dimensional offsets, one at the shoulder and the other at the elbow. 
We sampled eight two-dimensional offsets at the shoulder directionally 45 degrees from each other starting from the horizontal axis and at values with $30\%$ noticing probability. 
For each shoulder offset, we sampled the elbow joint two-dimensional offsets on circles shifted away from the origin opposite to the direction of the shoulder joint offset and by the amount of the shoulder offset value. 
The circles' radius range from $9$ to $24$ degrees with a $3$ degree interval and we sampled eight offsets spread evenly on each circle. 
We conducted data collection on four poses to validate the hypothesis. 
Each participant thus evaluated $8 \textit{shoulder offsets} \times 4 \textit{elbow offset sampling circles} \times 8 \textit{elbow offsets sampled per circle} \times 4 \textit{target poses}= 1024$ tasks. 
Each participant's data collection was divided into $8$ sessions with a five-minute break between sessions to avoid fatigue, lasting $90$ minutes in total. 

\subsubsection{Participants}
We recruited 12 participants including $8$ males and $4$ females. The participants' age ranged ranged from $19$ to $26$ $(AVG = 21.33,~SD = 1.28)$. All participants were right-handed. The self-reported familiarity score averages at $3.08$ (SD = $1.11$) with a 7-point Likert scale from $1$ (not at all familiar) to $7$ (very familiar).

\subsubsection{Apparatus}
The apparatus remains the same with Section \ref{section: apparatus}.

\subsubsection{Result}
\label{section:study1phase3result}
\begin{figure}[htbp]
    \begin{minipage}{0.42\textwidth}
        \centering
        \includegraphics[width=\linewidth]{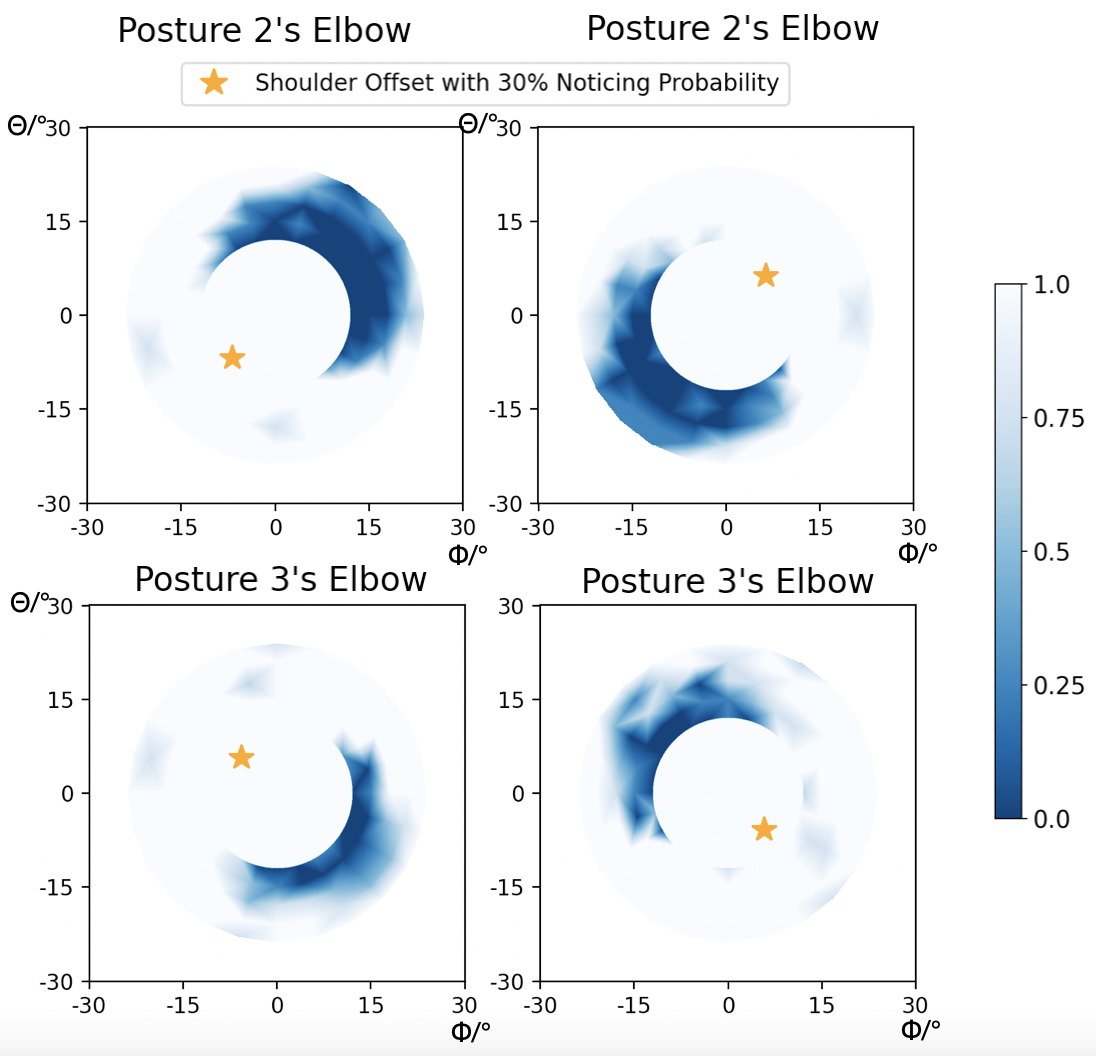}
        \caption{The noticing probability distribution of the composite offset with different shoulder offsets applied in the pilot study. The yellow stars label the applied shoulder offset. The darker color represents the lower noticing probability.}
        \label{fig:exp3pilot}
    \end{minipage}
    \hspace{10pt}
    \begin{minipage}{0.5\textwidth}
        \centering
        \includegraphics[width=\linewidth]{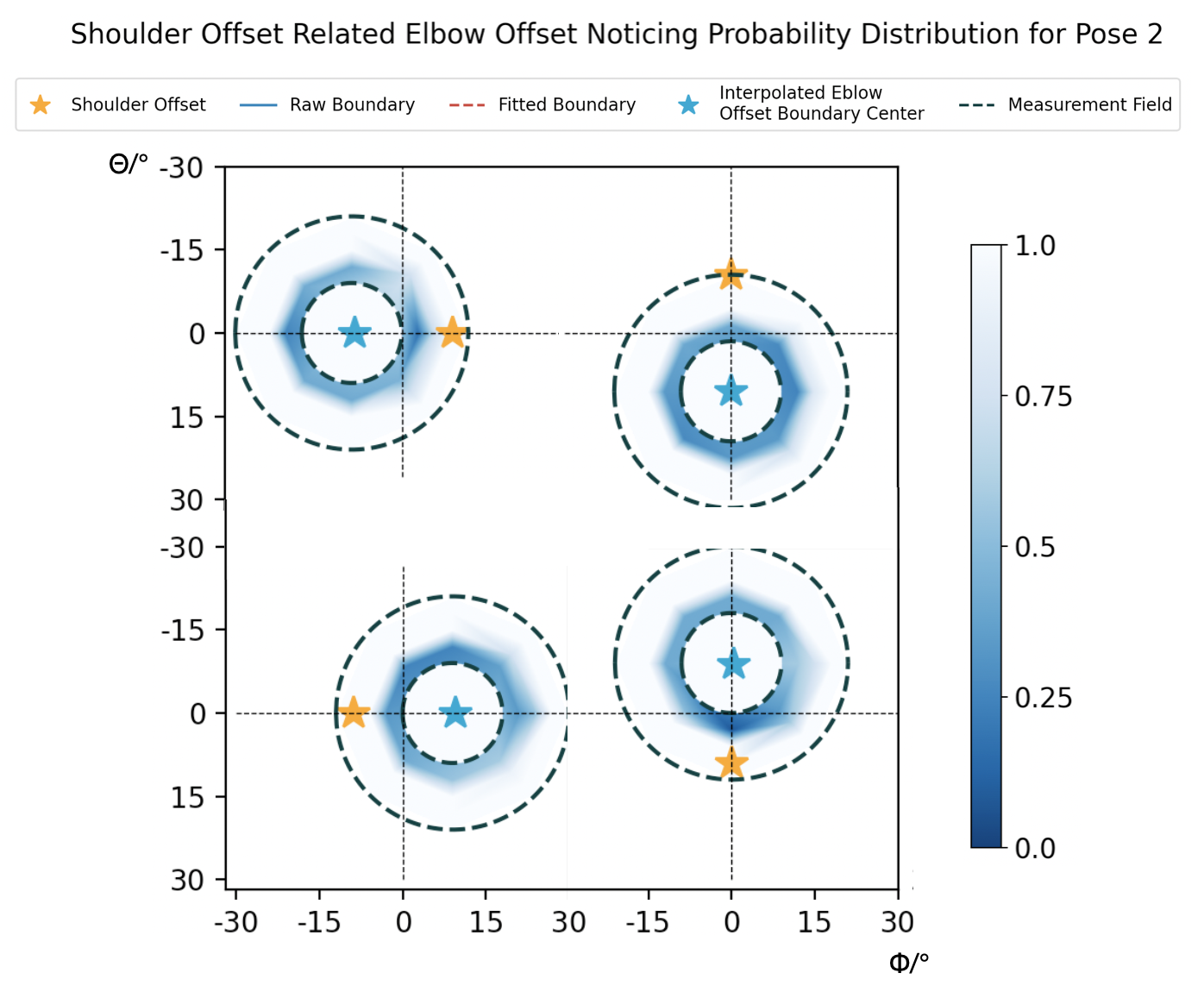}
        \caption{The noticing probability distributions of the composite offsets with different shoulder offsets applied at Pose 2. The heatmaps indicate the probability with darker color representing a lower noticing probability. The yellow stars label the applied shoulder joint offset and the blue stars are the interpolated elbow probability distribution center. The black dotted circles outline the task range.}
        \label{fig:exp3posture3}
    \end{minipage}
\end{figure}

Pose two examples of the elbow joint two-dimensional offset noticing probability affected by shoulder joint offsets of different directions are shown in Fig~\ref{fig:exp3posture3}. Refer to the supplementary material for a complete result. The blue stars are the center of the shifted elbow joint offset noticing probability distribution interpolated and the yellow stars show the applied shoulder joint offset. 
From the illustration, we found that the elbow joint offset oval shifts diagonally from the shoulder offset direction and for the same amount as the shoulder joint offset.
Thus for offsets $\phi_\text{s}$ and $\theta_\text{s}$ applied at the shoulder, the position of the shifted elbow joint probability distribution is centered at $(-\phi_\text{s},~-\theta_\text{s})$. We thus confirmed \textbf{H3} since \textbf{the probability distributions at two joints negatively and linearly affect each other when predicting the probability of noticing the offset with a given composite two-joint offset.} Therefore, given a composite two-joint offset, we can interpolate its probability of being noticed by the user.

\vspace{-3mm}
\section{Implementation of the offset model and the movement amplification technique}
\label{section:implementation}

Based on the results in Section~\ref{section:study1}, we quantify how the user-avatar movement inconsistency affects the sense of body ownership. Given an offset and a pose, we are able to interpolate the probability of noticing the offset. We obtain applicable offsets with a given acceptable offset noticing probability and a pose and adapt it to continuous movement in this section. Then we adapt it to continuous movement and implement an amplification technique as an instance of the model.

\subsection{Constructing the Model of Applicable Composite Two-joint Offsets}
\label{section:constructmodel}

\begin{figure}
    \centering
    \includegraphics[width=0.8\linewidth]{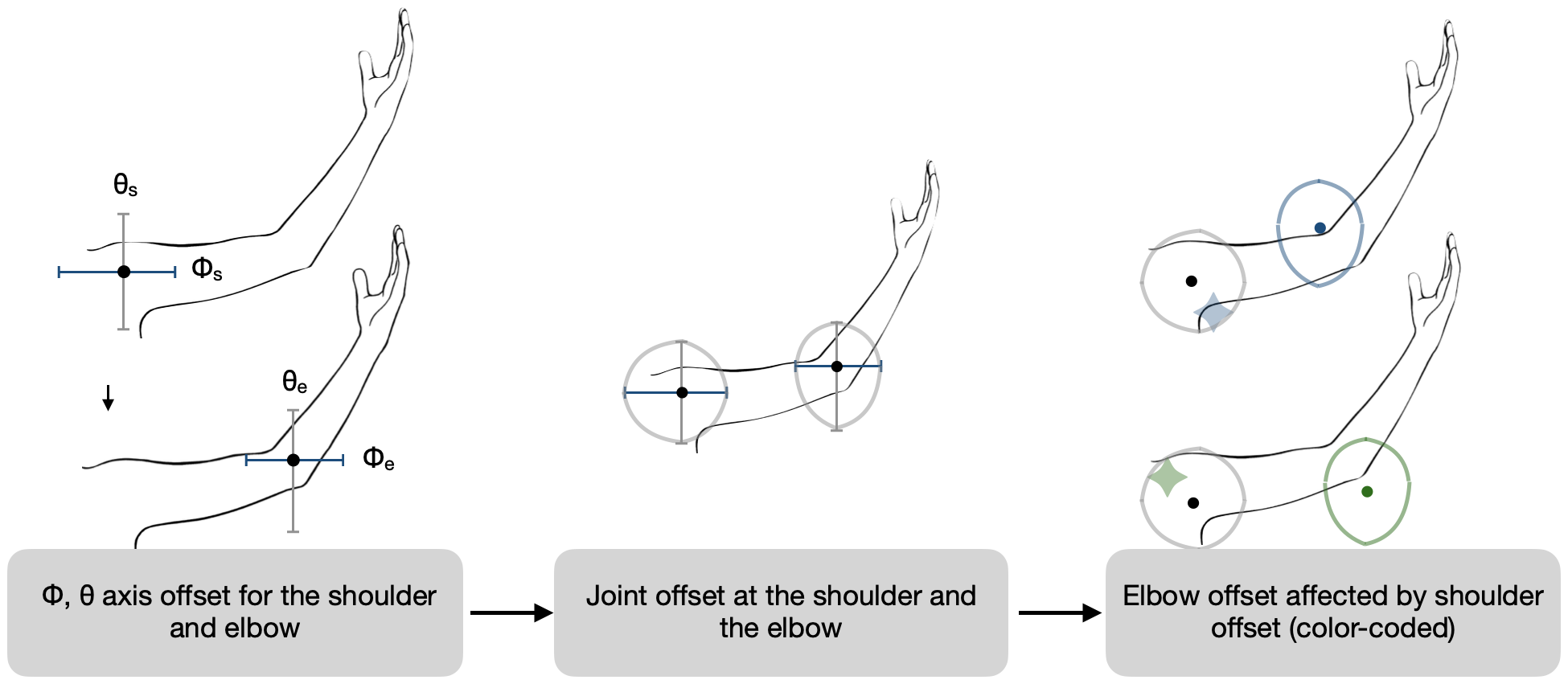}
    \caption{The three-step model of the offset with a given probability and pose. The left interpolates the single axis offsets with the given probability. The middle shows the two-dimensional offset with noticing probability equal to or less than the given probability on two joints separately. The right illustrates the composite two-joint offset with noticing probability equal to or less than the given probability.}
    \label{fig:pipeline}
\end{figure}

Leveraging the quantified relationship between the offset applied and the offset noticing probability, we construct a statistical model which outputs a set of applicable composite two-joint offsets given an offset noticing probability and a pose. 
With the explored offset noticing probability in Section~\ref{section:study1}, we build a statistical model to give out the offsets with a given noticing probability and pose.

Fig~\ref{fig:pipeline} outlines the model's pipeline. With an acceptable offset noticing probability \textit{p}, we first interpolate the single axis offset values at \textit{p} for all 10 poses and 4 axes based on the findings in Section~\ref{section:exp1}. 
Then for each axis, we fitted a relationship between the pose and the offset value at that axis based on which we can calculate the four single axis offset values ($\Phi_\text{s}$, $\Theta_\text{s}$, $\Phi_\text{e}$, and $\Theta_\text{e}$) at \textit{p} for any pose by Equation~\ref{equ:exp1}. 
A, B, C, D, and E are fitted parameters of the relationship. 

Then based on the characterized relationship between four single axis offsets ($\phi_\text{s}$, $\theta_\text{s}$, $\phi_\text{e}$, and $\theta_\text{e}$) and the two-dimensional offsets at a given probability for the shoulder and the elbow respectively in Equation~\ref{equ:ovalfunction}, we can interpolate the applicable oval-shaped two-dimensional offset for both joints with offset noticing probability \textit{p}.
Note that the applicable two-dimensional offsets are any point within the oval shape with offset noticing probability equal to or less than the required \textit{p}.

Section~\ref{section:exp3} defines the negative linear relationship between the elbow joint offset noticing probability and the shoulder joint offset noticing probability. 
The applicable composite two-joint offset is thus composed of the applicable shoulder joint offsets at \textit{p} and the accordingly shifted applicable elbow joint offsets at \textit{p} with a determined shoulder offset.

\begin{equation}
    \pm \{\phi, \theta\}_{\{s, e\}} = A \times \Phi_\text{s} + B \times \Theta_\text{s} +  C \times \Phi_\text{e} + D \times \Theta_\text{e} + E
    \label{equ:exp1}
\end{equation}

Algorithm \ref{alg:model} illustrates the algorithm procedure.

\algrenewcommand\algorithmicrequire{\textbf{Input:}}
\algrenewcommand\algorithmicensure{\textbf{Output:}}

\begin{algorithm}
\caption{Calculate the offsets of a specific noticing probability for an arbitrary pose}\label{alg:model}
\begin{algorithmic}[1]
\Require $\text{Given pose}(A_s, B_s, A_e, B_e)\text{, Given noticing probability }p, $
\Ensure $\text{A set of offsets of the given probability}\{(\alpha^p_s, \beta^p_s, \alpha^p_e, \beta^p_e)\}$
\State $(\phi^p_{si}, \theta^p_{si}, \phi^p_{ei}, \theta^p_{ei}) \gets f(\Phi_{si}, \Theta_{si}, \Phi_{ei}, \Theta_{ei})\text{ }i=1, 2, \ldots, 10$   
\hspace{1em}
\Comment{
$\phi^p_{si}$ is the $\phi$ offset on shoulder joint with probability $p$ on the $i$th sample pose. 
$f(\Phi_{si}, \Theta_{si}, \Phi_{ei}, \Theta_{ei})$ is the quadratic function fitted in Section~\ref{section:exp1} on the $i$th sample pose$(\Phi_{si}, \Theta_{si}, \Phi_{ei}, \Theta_{ei})$.}
\State $g(\Phi_s, \Theta_s, \Phi_e, \Theta_e) \gets g(\phi^p_{si}, \theta^p_{si}, \phi^p_{ei}, \theta^p_{ei}, \Phi_{si}, \Theta_{si}, \Phi_{ei}, \Theta_{ei})\text{ }i=1, 2, \ldots, 10$ 
\Comment{
$g(\Phi_s, \Theta_s, \Phi_e, \Theta_e)$ is the function fitted with Equation~\ref{equ:exp1} which takes in any pose $(\Phi_s, \Theta_s, \Phi_e, \Theta_e)$ and outputs the offset on four axes with the given probability $p$.}
\State $(\alpha^p_s, \beta^p_s, \alpha^p_e, \beta^p_e) \gets g(A_s, B_s, A_e, B_e)$  
\Comment{$(\alpha^p_s, \beta^p_s, \alpha^p_e, \beta^p_e)$ are the four single axis offsets with the given noticing probability $p$ on the given pose $(A_s, B_s, A_e, B_e)$.}
\State $F(\alpha_s, \beta_s) = P\{(\alpha_s, \beta_s)\text{'s noticing probability is equal to or less than p}\} \gets h(\alpha^p_s, \beta^p_s)$
\Comment{$F(\alpha_s, \beta_s)$ is the noticing probability distribution of the elbow offset. $h$ is the function defined by Equation~\ref{equ:ovalfunction}}.
\State $S_{s} = \{(\alpha_s, \beta_s) | F(\alpha_s, \beta_s) \leq 2p\}$  
\Comment{$S_s$ is a set of shoulder offsets ($\phi_s$, $\theta_s$) whose noticing probabilities $(\alpha_s, \beta_s)$ are equal to or less than twice of the given probability $p$.}
\State $F(\alpha_e, \beta_e) = P\{(\alpha_e, \beta_e)\text{'s noticing probability is equal to or less than p}\} \gets h(\alpha_e, \beta_e)$
\Comment{$F(\alpha_e, \beta_e)$ is the noticing probability distribution of the elbow offset. $h$ is the function defined by Equation~\ref{equ:ovalfunction}}.
\State $S_{e} = \{(\alpha_e, \beta_e) | F(\alpha_e, \beta_e) \leq 2p\}$
\Comment{$S_e$ is a set of elbow offsets ($\phi_e$, $\theta_e$) whose noticing probabilities $(\alpha_s, \beta_s)$ are equal to or less than twice of the given probability $p$.}
\State $F(\alpha_s, \beta_s, \alpha_e, \beta_e) = P\{(\alpha_s, \beta_s, \alpha_e, \beta_e)\text{'s noticing probability is equal to or less than p}\} \gets k(\alpha_s, \beta_s, \alpha_e, \beta_e)$
\Comment{$F(\alpha_s, \beta_s, \alpha_e, \beta_e)$ is the noticing probability distribution of the composite two-joint offset $(\alpha_s, \beta_s, \alpha_e, \beta_e)$. $k$ is the negative linear function based on Section~\ref{section:exp3} results.}
\State $S = \{(\alpha_s, \beta_s, \alpha_e, \beta_e) | F(\alpha_s, \beta_s, \alpha_e, \beta_e) \leq p, (\alpha_s, \beta_s) \in S_s, (\alpha_e, \beta_e) \in S_e\}$
\Comment{$S$ contains the composite offsets whose noticing probabilities are equal to or less than $p$.}
\end{algorithmic}
\label{alg:algorithm}
\end{algorithm}

\subsection{Implementation of the Dynamic Movement Amplification Technique}
\label{sec:implementation}

With a set of offsets of a specific probability, we could implement various functionalities by choosing different offsets. In this section, we implemented a dynamic amplification technique to demonstrate the usability and extendability of the model. 
Wentzel~\cite{wentzel2020improving} proposed a guideline for simulating such continuous movement which states that the modified motion should be continuous and should conform to ergonomics. A continuous movement can be defined with a series of poses. Sampling poses within the movement and calculating their maximum offsets is onerous and demanding. Furthermore, based on our model these static poses' offsets might vary and do not necessarily adhere to a monotonic relationship. Directly applying the maximum unnoticeable offset at each pose would result in incoherent virtual movements due to varying maximum unnoticeable offset of these poses. There are different solutions to simulate continuous movement smoothly and naturally based on varying optimization goals. 
We thus propose one such goal to continuously amplify movements and implement an interaction technique that amplifies the movement at the direction of the movement relative to the body. It involves finding an extreme pose based on the user's intentions-the extreme pose can be the target pose if a target exists or simply a directional extreme pose. The intended results are: offset increments linearly as the user moves towards the extreme pose; the maximum offset determined by the user defined maximum noticing probability is applied when the user reaches the extreme pose.

Here we describe in detail how we interpolate the four single axis offsets to be applied at any moment during the movement. Since we found that users could be less sensitive to the dynamic movement offsets, we adopted a $75\%$ offset noticing probability to determine the maximum offset. Parameterize the extreme pose by $\Phi_\text{S}, \Theta_\text{S}, \Phi_\text{E}$, and $\Theta_\text{E}$. At any given moment the current pose can be defined by $\Phi_\text{s}$, $\Theta_\text{s}$, $\Phi_\text{e}$, and $\Theta_\text{e}$. 
Given the extreme pose, we first obtain its two-dimensional shoulder joint maximum applicable offset. Since our technique applies offset along the direction of the pose, we interpolate the maximum extreme pose shoulder offset $\phi_{S}^{b}$ and $\theta_{S}^{b}$ at the pose shoulder's direction which implies Equation~\ref{eq:phithetashoulder}. We define the shoulder joint offset to be applied at the current pose linearly such that for each axis we have $\frac{\text{current offset}}{\text{current angle}} = \frac{\text{extreme offset}}{\text{extreme angle}}$. We can thus calculate the offsets to be applied at $\phi$ and $\theta$ axis from equation~\ref{eq:phishoulder}. From this we can deduct equation~\ref{eq:shoulder}. This confirms that the two-dimensional offset we applied at the shoulder conforms with the current shoulder's direction relative to the origin.

\begin{equation}
    \frac{\phi^{b}_\text{S}}{\theta^{b}_\text{S}} = \frac{\Phi_\text{S}}{\Theta_\text{S}}
    \label{eq:phithetashoulder}
\end{equation}

\begin{equation}
   \phi_\text{s} = \frac{\Phi_\text{s}}{\Phi_\text{S}} \times \phi^{b}_\text{S}
    ,
    \theta_\text{s} = \frac{\Theta_\text{s}}{\Theta_\text{S}} \times \theta^{b}_\text{S}
    \label{eq:phishoulder}
\end{equation}

\begin{equation}
    \frac{\phi_\text{s}}{\theta_\text{s}} = \frac
    {\frac{\Phi_\text{s}}{\Phi_\text{S}} \times \phi^{b}_\text{S}}
    {\frac{\Theta_\text{s}}{\Theta_\text{S}} \times \theta^{b}_\text{S}} = \frac
    {\frac{\Phi_\text{s}}{\Phi_\text{S}} \times \Phi_\text{S}}
    {\frac{\Theta_\text{s}}{\Theta_\text{S}} \times \Theta_\text{S}} = 
    \frac{\Phi_\text{s}}{\Theta_\text{s}}
    \label{eq:shoulder}
\end{equation}

\begin{figure}[htbp]
    \centering
    \includegraphics[width=0.4\linewidth]{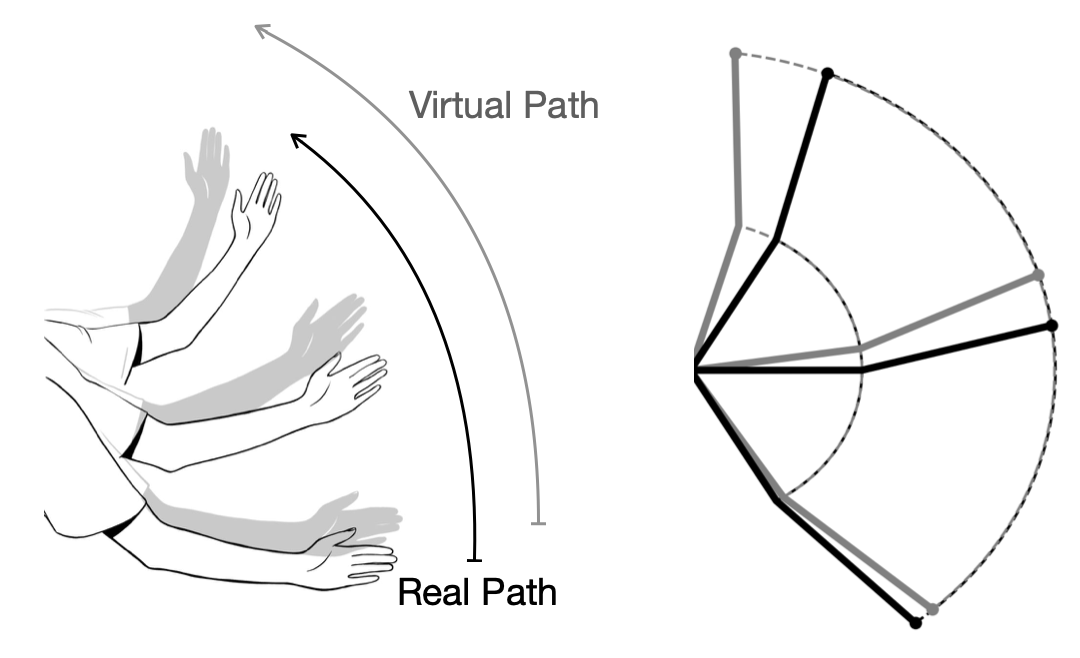}
    \caption{The physical and virtual arms are shown in black and grey respectively. The shoulder and elbow paths are illustrated by dotted lines. The offset increases when the physical arm is raising.}
    \label{fig:dynamic}
\end{figure}

With the determined shoulder joint offset, we can use the model to obtain the applicable elbow joint offsets. The above steps can then be repeated to choose the applicable elbow joint offset. 
Fig~\ref{fig:dynamic} shows an example of movement amplification by our technique. The left shows the amplified movement as the user raises his arm upwards and the right illustrates the programmed tracking of the physical and virtual arm movement paths. 

\section{Evaluating the Amplification Technique}
\label{section:study2}

In Section \ref{sec:implementation}, we implemented an interaction technique that amplifies the user's arm movements by adding dynamic unnoticeable angular offsets to their shoulder and elbow.
To validate that technique achieves the design goal and evaluate the user performance, we conduct a comparison experiment with the task of performing different arm poses in VR.
We compared five implementations of the proposed techniques with the difference in two key parameters (offset strength, offset distribution).
We calculated task completion time, physical, and RULA score~\cite{mcatamney1993rula} as the quantitative metrics and collected the participants' subjective feedback in post-experiment questionnaires.
Results showed that adding offsets significantly reduced physical efforts, a strength of medium level helped participants maintain body ownership, and a balanced offset on two joints was generally favored by the participants.

\vspace{-3mm}
\subsection{Task}
\label{section:evatask}
The task for the participant was to control the virtual avatar's arm with their own arm movements to perform the target arm poses.
We rendered a virtual avatar whose viewpoint coincides with that of the users in VR.
The avatar's arm moved as the user's physical arm, whose shoulder, elbow, and hand position was tracked with a motion capture system.
In four of the five test conditions, we added dynamic angular offsets calculated with the proposed model to the shoulder and the elbow of the virtual avatar while the arm was moving.
To indicate the target pose, we rendered two blue balls in mid-air, showing the target positions of the elbow and the wrist joint positions, respectively.
The participants needed to overlap the two balls with the elbow and the wrist of the virtual avatar.
Visual feedback of turning the balls to the color green was provided when the participant managed to overlap them with the correct joints.
Time duration of $0.5$ seconds keeps the target pose triggered the completion of the task.
In between the tasks, the participant reset their arm poses to naturally lay down on the side.
We asked the participants to keep the remaining part of the body still while changing the arm pose during the tasks.
Similar to Section~\ref{section: clusterposes}, we used the HDBSCAN cluster algorithm on the CMU MoCap dataset to sample $40$ target poses. 
Considering the fatigue might affect the results, participants do not need to repeat the same pose in one condition.
Each participant thus completed $5 \textit{conditions} \times 40 \textit{poses} = 200$ tasks.
We used a latin square to counterbalance the order of condition and the tasks were randomized.

\vspace{-3mm}
\subsection{Design}
\label{section:evadesign}
We used a within-subject factorial design with the independent variables of \emph{offset strength} (no, medium, strong) and \emph{offset distribution} (shoulder, shoulder and elbow, elbow).
\emph{Offset strength} was to test whether adding unnoticeable offsets (no V.S. medium) can improve user performance and maintain body ownership at the same time, and the trade-off of increasing the performance gain and losing body ownership (medium V.S. strong).
We controlled the offset strength with the predicted noticeability of the offsets at the extreme arm pose $(90,~90,~90,~90)$, in which the participant should be the most sensitive about the offset.
We set the noticing probability to be $75\%$ and $100\%$ for the medium and strong levels, respectively.
Note that as predicted by the proposed model, we expected the offsets at the strong level to be possibly noticed by the participants.
\emph{Offset Distribution} was to compared adding composite two-joint offsets to the naive conditions when it degrades to adding one-joint offsets to either the shoulder or the elbow.
Specifically, we scaled its shoulder offset by $\delta_\text{s}$ and its elbow offset by $\delta_\text{e} = 1 - \delta_\text{s}$. 

\begin{table}[htbp]
\caption{Offset strength and offset distribution parameters for the conditions.}
  \begin{tabular}{ccccc}
    \toprule
    Condition & Noticing probability & $\delta_\text{shoulder}$ & $\delta_\text{elbow}$ & Description \\
    \midrule
    \textit{N} & $0\%$ & N/A & N/A & No offset \\
    \textit{MB} & $75\%$ & $0.5$ & $0.5$ & Medium level, balanced offset \\
    \textit{ME} & $75\%$ & $0$ & $1$ & Medium level, only elbow offset \\
    \textit{MS} & $75\%$ & $1$ & $0$ & Medium level, only shoulder offset \\
    \textit{HB} & $100\%$ & $0.5$ & $0.5$ & High level, balanced offset \\
  \bottomrule
\end{tabular}
\label{tab: freq}
\end{table}

As shown in Table \ref{tab: freq}, we compared five implementations in total.
In comparing conditions \textit{N}, \textit{MB}, and \textit{HB}, we investigated the level of offset strength; while
in comparing conditions \textit{MB}, \textit{ME}, and \textit{MS}, we investigated the distribution of offset at two joints. 
We measured the completion time, Rapid Upper Limb Assessment(RULA) score~\cite{mcatamney1993rula}, the virtual and physical path length for the elbow and wrist as quantitative metrics for task performance.
Task completion time is the time taken by the participant to reach the target arm pose from the reset pose. 
RULA score measures the ergonomic difficulty of a pose. 
It is an objective metric which can be calculated with the shoulder and elbow angles.
A pose can be graded with a series of rules, for instance, if the upper arm puts downward, the RULA score will plus one, while if the upper arm raises upward, the RULA score will plus four.
A lower RULA score indicates a lower physical effort and a more comfortable pose. 
The virtual path length refers to the ratio of the avatar movement path length to the linear distance between the initial and the target pose. 
Similarly, physical path length refers to the user's real movement path length to target distance ratio. 
Then after each condition session, each participant was asked to fill a survey to report their sense of comfort, ease of reach, sense of control, and body ownership based on a 7-point Likert scale (1: strongly disagree, 4:neutral, 7: strongly agree).
All user studies had been approved by our university's IRB board.

\vspace{-3mm}
\subsection{Apparatus and participants}

We developed an experimental platform with Unity 2021 for Oculus Quest 2 headset. The interface showed the virtual avatar as well as the blue balls indicating the target arm pose. 
We captured the participant's movement with Optitrack and applied our dynamic amplification technique on the virtual arm. Then the modified movement data was sent to the headset and was rendered to the user at $30$ FPS. The system's hardware and setup remained the same with Section \ref{section: apparatus}.

Given that the most related works~\cite{li2021armstrong, tiare2018ownershift} conducted their evaluation studies with 12 users, we recruited 12 participants for the evaluation. 
All of them had not participated in our previous experiments and were right-handed. 
The participants aged from 19 to 22 $(AVG = 20.25,~SD = 0.63)$, including 5 females and 7 males. 
Half of the participants had at least moderate experience with virtual reality. 
The average self-reported virtual reality familiarity was 2.58 (1: Not familiar at all; 7: Very familiar).

\vspace{-3mm}
\subsection{Results}
\label{section:evaresults}
We conducted the Bartlett's test and validated that the assumption of sphericity had not been violated on the metrics of the task completion time$(\chi^2 = 2.09,~p > 0.05)$, RULA score$(\chi^2 = 0.14,~p > 0.05)$, physical elbow$(\chi^2 = 8.78,~p > 0.05)$, physical wrist$(\chi^2 = 7.93,~p > 0.05)$.
Therefore, we conducted repeated measures ANOVA tests on these metrics.
If any independent variable had significant effects$(p < 0.05)$, we used Bonferroni-corrected post hoc T-tests for pairwise comparisons since we compared multiple pairs of data.
Since the subjective scores were discrete and ordinal, they were analyzed using Friedman tests with a series of Wilcoxon signed-rank post hoc tests. 
We performed our data analysis using scipy~\cite{scipy}.
We compare conditions \textit{N}, \textit{MB}, \textit{HB}, and \textit{MB}, \textit{ME}, \textit{MS} to study the effect of \emph{offset strength} and \emph{offset distribution} respectively.

RM-ANOVA tests showed significant differences on task completion time $(F_{4, 44} = 31.94,~p < 0.05,~\eta^2 = 0.74)$, RULA score$(F_{4, 44} = 16.59,~p < 0.05,~\eta^2 = 0.60)$, physical moving distance of elbow$(F_{4, 44} = 10.03,~p < 0.05,~\eta^2 = 0.60)$, and physical moving distance of wrist$(F_{4, 44} = 7.48,~p < 0.05,~\eta^2 = 0.51)$.
% but not on virtual moving distance of elbow$(F_{4, 44} = 1.34, p > 0.05)$ and wrist$(F_{4, 44} = 0.87, p > 0.05)$.
We confirmed that the virtual path lengths of all participants had no significant differences to ensure that they followed the same target path in all conditions and the offset would not result in extra movements.

\subsubsection{Effects of offset strength.}

\begin{figure}[htbp]
    \centering
    \includegraphics[width=\linewidth]{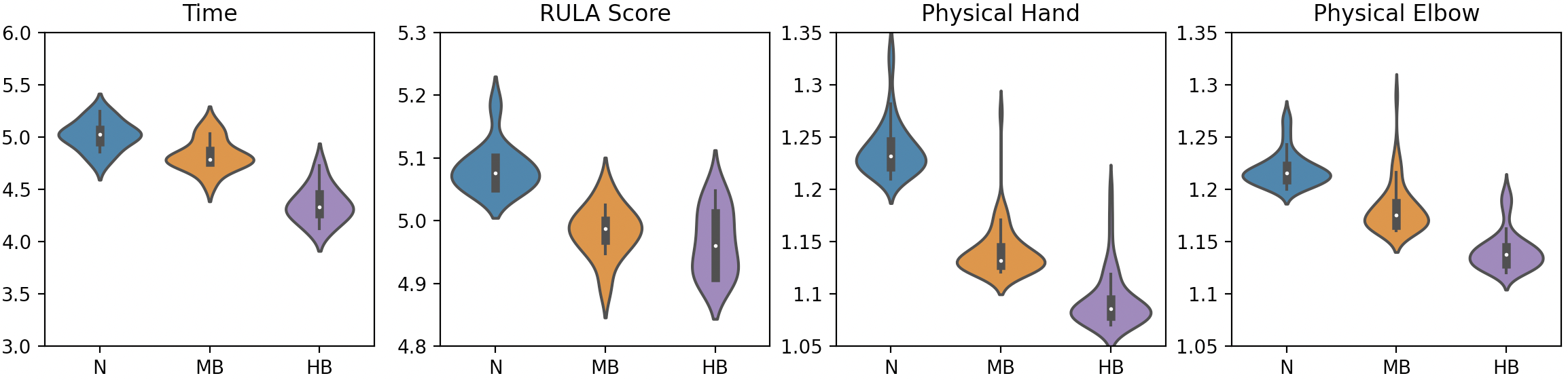}
    \caption{Quantitative results of the three techniques with different offset strength parameters. }
     \label{fig:study2abe_quant}
\end{figure}

\textit{Adding offsets significantly reduced task completion time.}
Pair-wise comparisons showed significant differences on task completion time between the following pairs: 
\textit{N} $(AVG = 5.02,~SD = 0.13)$ and \textit{MB} $(AVG = 4.83,~SD = 0.14,~t(11) = 2.88,~p < 0.05,~d = 1.29)$, 
\textit{MB} $(AVG = 4.83,~SD = 0.14)$ and \textit{HB} $(AVG = 4.35,~SD = 0.16,~t(11) = 7.42,~p < 0.05,~d = 2.94)$, 
\textit{N} $(AVG = 5.02,~SD = 0.13)$ and \textit{HB} $(AVG = 4.35,~SD = 0.16,~t(11) = 12.56,~p < 0.05,~d = 4.21)$.

\textit{Adding offsets reduced the physical effort and made movement ergonomically easier.} 
Pair-wise comparisons showed significant differences on the RULA score between 
\textit{N} $(AVG = 5.08,~SD = 0.04)$ and \textit{MB} $(AVG = 4.98,~SD = 0.04,~t(11) = 6.29,~p < 0.05,~d = 2.57)$, 
\textit{N} $(AVG=5.08, SD=0.04)$ and \textit{HB} $(AVG = 4.97,~SD = 0.05,~t(11) = 5.78,~p < 0.05,~d = 2.36)$.

\textit{Adding offsets effectively reduced the physical moving distance, making the tasks less physically demanding.} 
The effect of larger offsets on decreasing physical path length was significant on both elbow.
Pair-wise comparisons showed significant differences on physical path length between all pairs: 
\emph{elbow} \textit{N} $(AVG = 1.22,~SD = 0.02)$ and \textit{MB} $(AVG = 1.19,~SD = 0.03,~t(11) = 3.17,~p < 0.05,~d = 1.03)$, 
\textit{MB} $(AVG = 1.19,~SD = 0.03)$ and \textit{HB} $(AVG = 1.14,~SD = 0.02,~t(11) = 3.18,~p < 0.05,~d = 1.73)$, 
\textit{N} $(AVG = 1.22,~SD = 0.02)$ and \textit{HB} $(AVG = 1.14,~SD = 0.02,~t(11) = 3.51,~p < 0.05,~d = 3.43)$; 
\emph{wrist} \textit{N} $(AVG = 1.24,~SD = 0.03)$ and \textit{MB} $(AVG = 1.15,~SD = 0.03,~t(11) = 2.23,~p < 0.05,~d = 1.54)$, 
\textit{MB} $(AVG = 1.15,~SD = 0.03)$ and \textit{HB} $(AVG = 1.10,~SD = 0.03,~t(11) = 3.16,~p < 0.05,~d = 1.29)$, 
\textit{N}$(AVG = 1.24,~SD = 0.03)$ and \textit{HB} $(AVG = 1.10,~SD = 0.03,~t(11) = 4.75,~p < 0.05,~d = 2.32)$

\begin{figure}[htbp]
    \centering
    \includegraphics[width=\linewidth]{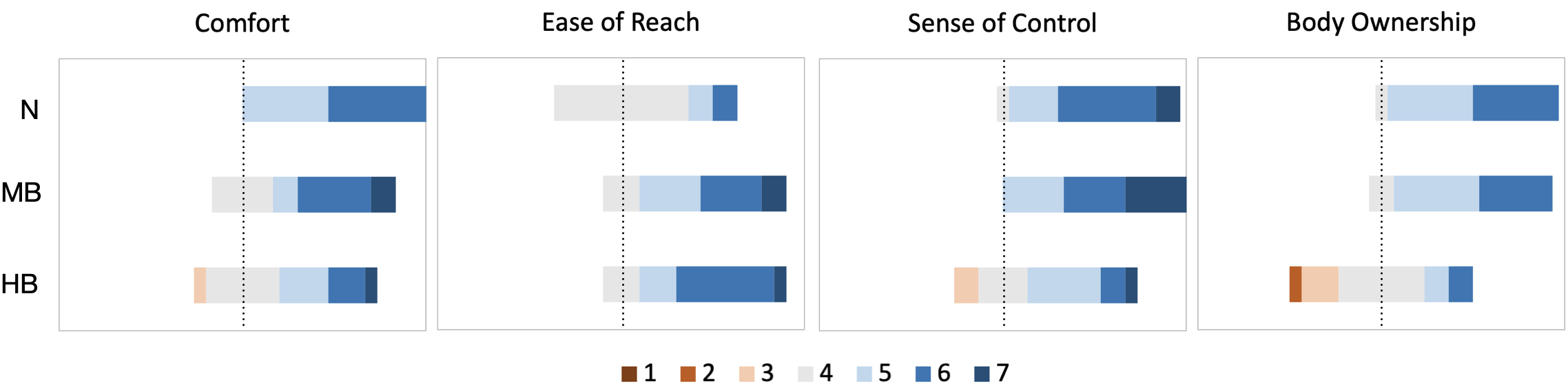}
    \caption{Qualitative results of the three techniques of different offset strength parameters.}
    \label{fig:study2subabe}
\end{figure}

\textit{Strong offsets reduced the comfort and body ownership.} The Wilcoxon signed-rank test results show that comfort, sense of control, and body ownership is only reduced with significance when offsets of strong strength were applied. 
For the comfort metric, large offsets reduced comfort as \textit{N} $(AVG = 5.53,~SD = 0.50)$ was significantly higher than \textit{HB} $(AVG = 4.80,~SD = 1.05,~Z = 2.09,~p < 0.05)$.
Body ownership was reduced significantly in \textit{HB} $(AVG = 4.07,~SD = 1.06)$ compared to \textit{N}$(AVG = 5.40,~SD = 0.61)$ and \textit{MB}$(AVG = 5.27,~SD = 0.68)$: \textit{N} and \textit{HB} $(Z = 2.83,~p < 0.05)$, \textit{MB} and \textit{HB} $(Z = 2.77,~p < 0.05)$. 
\textit{Users found it hard to control the virtual arm with a strong offset.} Sense of control is reduced significantly in \textit{HB} $(AVG = 4.73,~SD = 1.06)$ compared to both \textit{N} $(AVG = 5.73,~SD = 0.77)$ and \textit{MB} $(AVG = 6.04,~SD = 0.82)$: \textit{N} and \textit{HB} $(Z = 2.76,~p < 0.05)$, \textit{MB} and \textit{HB} $(Z = 2.82,~p < 0.05)$.
\textit{Users felt easier to move and reach with added offsets.}
Ease of reach improved for both \textit{MB} $(AVG = 5.40,~SD = 0.95)$ and \textit{HB} $(AVG = 5.47,~SD = 0.88)$ compared to \textit{N} $(AVG = 4.40,~SD = 0.71)$ with significance: \textit{N} and \textit{HB} $(Z = -2.97,~p < 0.05)$, \textit{N} and \textit{MB} $(Z = -2.64,~p < 0.05)$. This implies the applying offset made reaching tasks easier yet such effect isn't raised when higher level offset is applied. 

The quantitative result demonstrates that only a strong offset reduced the ergonomic difficulty of the tasks. With an increasing level of offsets applied, users take less time and move less to achieve ergonomically easier tasks. 
The qualitative result also supported that the ease of reach improves with offsets and supplemented that users' comfort, sense of control, and body ownership are only compromised with significance when a high-level offset is applied. This reflects that virtual tasks can be made easier, more efficient, and more accessible without compromising body ownership illusion by applying a medium level of offsets. 

\subsubsection{Effects of offset distribution}

\begin{figure}
    \centering
    \includegraphics[width=\linewidth]{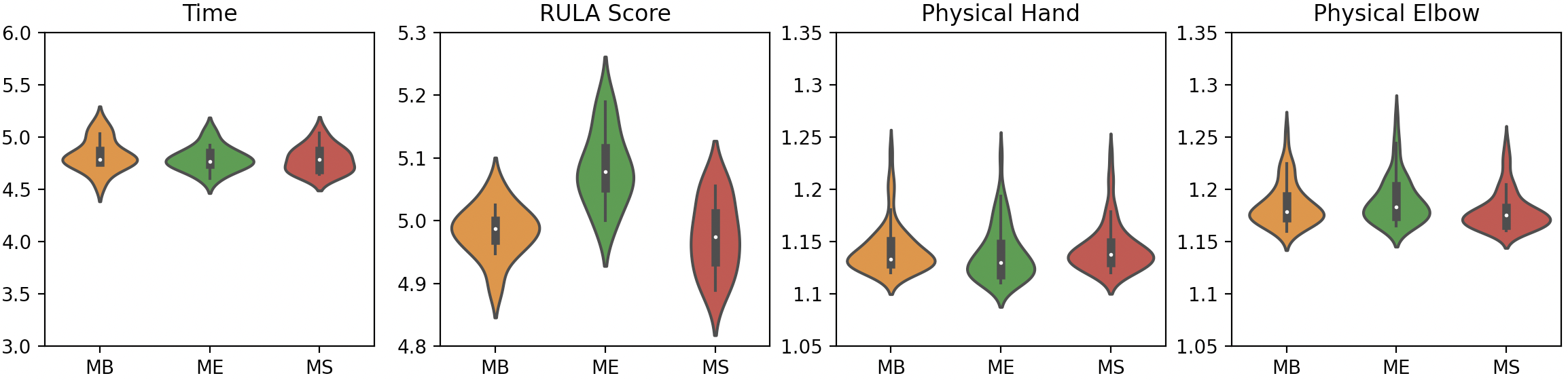}
    \caption{Quantitative results of the three techniques of different offset strength parameters.}
    \label{fig:study2bcd_quant}
\end{figure}

\textit{The balanced distribution of offset reduced the physical effort most.} 
Pair-wise comparisons showed significant differences only on RULA score.
RULA score is the highest with unbalanced offset when only the elbow has offset applied as \textit{ME} $(AVG = 5.08,~SD = 0.06)$ is significantly higher than both \textit{MB} $(AVG = 4.98,~SD = 0.04,~t(11) = -5.05,~p < 0.05,~d = 1.97)$ and \textit{MS} $(AVG = 4.97,~SD = 0.06,~t(11) = 4.72,~p < 0.05,~d = 1.83)$. This implies that compared to shoulder offset, elbow offset adds more to ergonomic unease. None other condition or metric is significant. 

\begin{figure}[htbp]
    \centering
    \includegraphics[width=\linewidth]{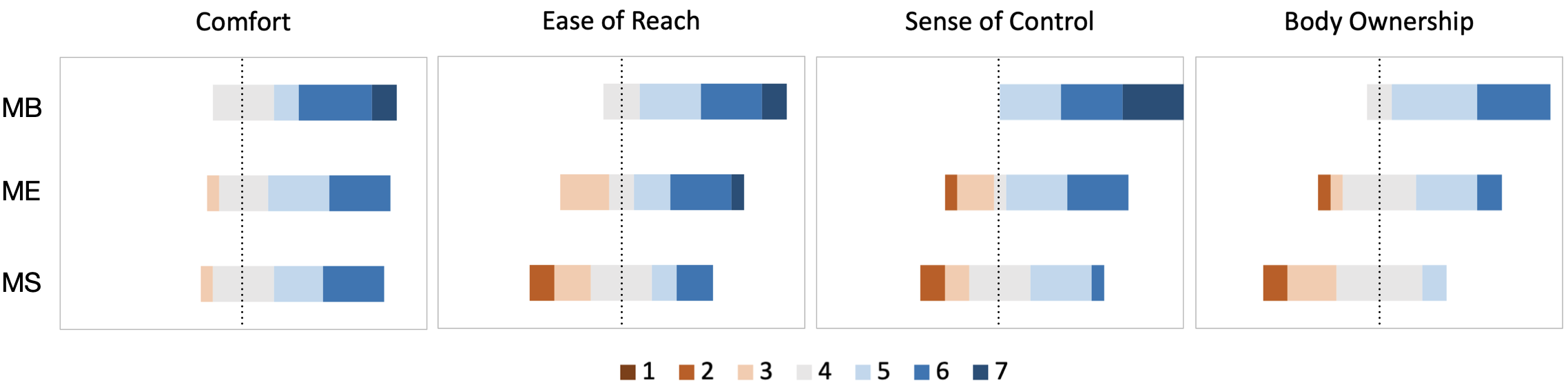}
    \caption{Qualitative results of the three techniques of different offset strength parameters.}
    \label{fig:study2subbcd}
\end{figure}

\textit{Unbalanced distribution of offset made the user feel hard to control and reach.} For the ease of reach metric, \textit{MB} $(AVG = 5.40,~SD = 0.95)$ is higher than both \textit{ME} $(AVG = 4.80,~SD = 1.33)$ and \textit{MS} $(AVG = 4.07,~SD = 1.29)$ with significance: \textit{MB} and \textit{ME} $(Z = 2.08,~p < 0.01)$, \textit{MB} and \textit{MS} $(Z = 2.87,~p < 0.01)$. Similarly, applying balanced offset achieves the best sense of control for users as \textit{MB} $(AVG = 6.04,~SD=0.82)$ significantly outperforms both \textit{ME} $(AVG = 4.67,~SD = 1.30)$ and \textit{MS} $(AVG = 4.07,~SD = 1.12)$: \textit{MB} and \textit{ME} $(Z = 2.92,~p < 0.01)$, \textit{MB} and \textit{MS} $(Z = 4,01,~p < 0.01)$. 

\textit{The balanced distribution of offset made the virtual movement more natural.} The body ownership subjective rating is the highest for balanced offset while within two unbalanced offset distribution, applying elbow offset has higher body ownership. \textit{MB} $(AVG = 5.27,~SD = 0.68)$ scores higher than \textit{ME} $(AVG = 4.42,~SD = 1.02)$ which outperforms \textit{MS} $(AVG = 3.63,~SD = 0.88)$: \textit{MB} and \textit{ME} $(Z = 2.69,~p < 0.01)$, \textit{ME} and \textit{MS} $(Z = 2.15,~p < 0.05)$, \textit{MB} and \textit{MS} $(Z = 3.31,~p < 0.01)$. Several participants reported that they felt the virtual arm not part of their body since the unbalanced offset made the virtual arm movement unnatural and thus hurt the body ownership.

The quantitative result demonstrates that the completion time and physical movement path ratio for both joints are unaffected by offset distribution. On the other hand, the RULA score indicates that the task is more ergonomically difficult when the offset is only applied at the elbow. This implies that while offset distribution doesn't affect the efficiency of the task completion in VR, offset unevenly applied on the elbow is less effective in making the tasks ergonomically easier thus more comfortable to achieve. Users' subjective preferences for offset distribution, however, are more significant. Balanced distribution helps improve ease of reach, sense of control, and the sense of body ownership. Furthermore, body ownership is reduced the most with an unbalanced offset applied only at the shoulder. Comfort is unaffected by offset distribution.

\vspace{-2mm}
\section{Applications}
\label{section:application}
To demonstrate our model's potential in leveraging the trade-off between functionality and body ownership illusion, we developed three applications with different functionality versus body ownership needs.

\vspace{-2mm}
\subsection{Motivating VR Stroke Rehabilitation}

Therapeutically appropriate rehabilitation is crucial to post-surgery motor function recovery for stroke patients~\cite{kilgard1998cortical, sathian2011neurological, sawaki2008constraint, teasell2005s}. 
We propose to provide virtual motor performance improvement to motivate patients and thus to boost engagement\cite{kwakkel2006impact, hee2020rehab, radomski2011more} in rehabilitation exercises.
In order to enable the patients to embody the virtual avatar, VR rehabilitation has a high requirement for body ownership.
We thus developed a stroke VR rehabilitation system that motivates the patients through subtle motion amplification. 
In a living room environment, the patient is instructed to perform stroke rehabilitation poses (abduction-adduction, internal-external rotation, elbow flexion-extension, and forearm pronation-supination) by coinciding the mid-air balls which outline the shoulder and the elbow joint trajectories.
During this process, the virtual left arm position can be subtly amplified as compared to the real physical arm by applying a small offset. 
In perceiving a motor performance improvement virtually, the patient gains stronger engagement and will likely to be motivated and spend more effort in rehabilitation. 

\vspace{-2mm}
\subsection{Manipulating Physical Effort Demand in VR Action Games}
\begin{figure}[htbp]
    \centering
    \begin{subfigure}{0.3\textwidth}
        \centering
        \includegraphics[width=\linewidth]{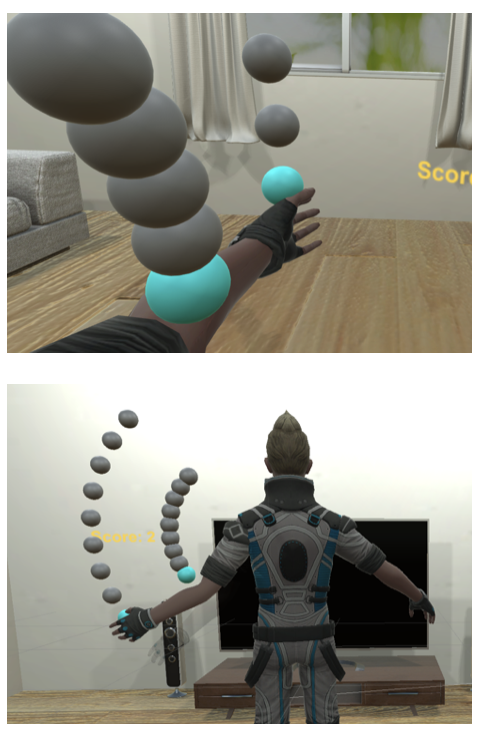}
        \caption{ }
        \label{fig:apprehab}
    \end{subfigure}
    \hfill
    \begin{subfigure}{0.3\textwidth}
        \centering
        \includegraphics[width=\linewidth]{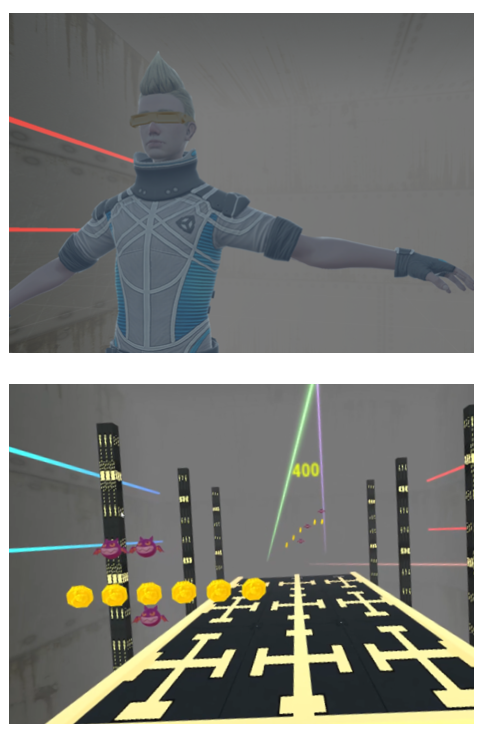}
        \caption{ }
        \label{fig:apppose}
    \end{subfigure}
    \hfill
    \begin{subfigure}{0.3\textwidth}
        \centering
        \includegraphics[width=\linewidth]{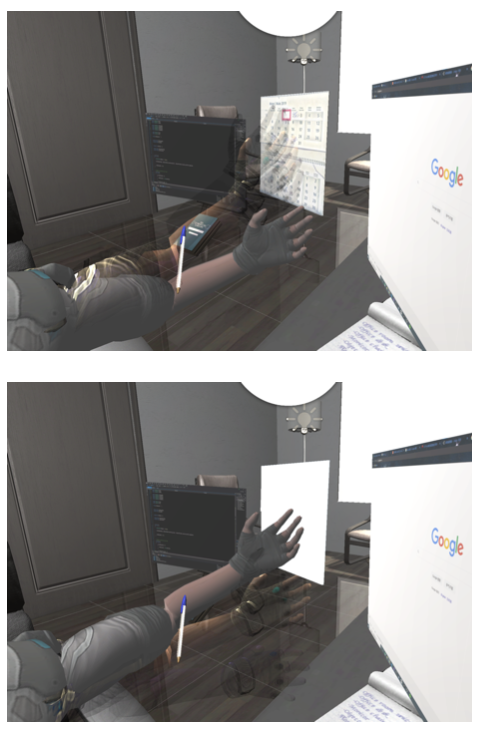}
        \caption{ }
        \label{fig:appwidget}
    \end{subfigure}
    \caption{The illustration of three applications. (a) Motivating stroke rehabilitation: The mid-air balls guide the user's arm for rehabilitation. (b) Changing physical effort demand in VR action games: The mid-aid balls illustrates the target pose. The level of offset applied onto the arm can be adjusted to change the game's demand for physical effort. (c) Input augmentation for widget arrangement: Users can amplify their arm movement to different extents by applying varying levels of offsets.}
    \label{fig:applications}
\end{figure}

With the model we propose motion modification as a new dimension in manipulating users' physical effort demand~\cite{eyesfree, headgesture} in VR games and therefore simulating a virtual avatar's stamina level. Specifically, the physical effort demand can be lowered by motion amplification and increased by motion reduction.
We thus designed a first-person VR action game inspired by VR games Beat Saber~\cite{beatsaber} and Oh Shape~\cite{ohshape}. 
The players need to pose their left arm to coincide with the target pose, represented by mid-air coins, and to avoid colliding with the devil bats.
Example scenes of the game are shown in Fig~\ref{fig:applications}. 
The player's motion is amplified with a medium-level offset, so the virtual arm moves quicker. 
Similarly the virtual arm moves slower when the same level of offset is applied opposite to the moving direction.

\vspace{-2mm}
\subsection{Widget Arrangement in VR}

Movement in a large space has been unpopular in VR both due to long term fatigue induced by large movements, or limited physical movement space. 
Input augmentation techniques can help transform small-scale and close-to-body movements in the real world to large-scale and more dramatic virtual motions. 
Such functionality oftentimes comes at the expense of body ownership illusion.
We developed an demonstration for widget arrangement tasks~\cite{semanticAdapt} based on our model with high functionality requirements yet relatively low body ownership requirement. 
In a typical work/study scenario, when the user is sitting at a desk and has limited space in the real world, users can reach distant positions with shortened moving distance, less time, and less physical energy with amplified movements.

\vspace{-2mm}
\section{Discussion}

In this paper, we have thoroughly investigated the effects of user-avatar movement inconsistency on its noticeability. 
We constructed a model that predicts a set of applicable offsets to an arm pose for a required noticing probability. 
With this model, we implemented an interaction technique that amplifies the user's arm movement without letting them notice the offsets. 
We developed three example applications to demonstrate the extendability of the proposed method. 
In this section we discuss the limitations of our work and suggest new opportunities for future work.

\vspace{-2mm}
\subsection{Effects of Avatar's Appearance on Body Ownership}
\label{section:discussavatar}
In this paper, we used the default humanoid avatar in Unity 2021 to study how the offset affects body ownership. 
Since the avatar's appearance could affect body ownership~\cite{Yee2007, Kilteni2012, dongsik2017theimpact}, we fixed the avatar to avoid the effects of this confounding factor. 
Since the changes in the avatar's appearance have reduced the sense of body ownership, the user might be more sensitive to the motion offset. However, we acknowledge the potential of modifying the avatar's appearance and the motion at the same time to provide interaction functionalities and to keep the body ownership as well, which will be important future work of this paper.

\vspace{-2mm}
\subsection{Applying Offsets beyond Angular Rotations on Shoulder and Elbow}
The offsets we studied in this paper are the angular rotation offsets applied on the shoulder and the elbow joints.
We acknowledge that there are offsets of other types that can also be applied to modify the movements.
Existing research~\cite{wentzel2020improving, ivan1996go, mcintosh2020iteratively} has explored stretching the arm without a length restriction to enable reaching the space far away.
We expect the research approach that we adopted in this study can also be applicable to the investigation of other offsets, including length and the combined usage of length and rotation offsets.

On the other hand, we did not investigate the noticeability of hand and finger movement inconsistency in this paper, considering the relatively limited influence imposed by hand and finger to the entire arm pose.
However, as one of the most frequently used organs, hand and finger could play an important role in the influences of user-avatar movement inconsistency in VR.
Though we did not involve hand and finger in our study, we expect that the research flow in our study can be adopted to hand movement inconsistency, finger, and composite inconsistency in the future.

\vspace{-2mm}
\subsection{Calculate the Noticing Probability for a Given Offset and Pose}

In Section~\ref{sec:implementation}, we constructed a model which took in a given probability and arm pose and predicted a set of applicable offsets. This provides convenience for application developers with an explicit requirement on the noticeability of the movement inconsistency, and the model automatically calculates the applicable offsets for them to adopt. However, we could also construct the model reversely as calculating the noticing probability of the movement inconsistency given a pair of offsets to apply on the shoulder and the elbow joints. With the collected noticing probability distributions of the offset strength on $10$ poses, the model can interpolate the offset strength and the arm pose to calculate the probability for a given pose and offset. This could help other researchers or engineers who would like to know how many users could notice the offset in their projects or research. 

\subsection{Human Adaptation to the Offset}
\label{section:discussadapt}
Though we gave out a model to tell the noticeability of a certain offset and pose, users might adapt to the offset if it continuously exists. 
This could change the noticeability to the offset which indicate that our model might add the adaptation as a parameter.
After our user studies, we divided every user's data into four groups in chronological order.
Then we conducted repeated measures ANOVA tests with the variables of chronological order on the number of users reported that they noticed the offset after checking the sphericity had not been violated.
The results showed that the task order did not significantly affect the noticing times $(F_{3, 31} = 1.83,~p > 0.05)$ which indicated that users did not adapt to the offset in $90$ minutes experiment.
However, if long term offset was applied, users could be less sensitive to the offset since they believed that some modified movements were consistent with physical movements and became more acceptive to larger offsets.

\vspace{-2mm}
\subsection{Limitation and Future Work}
We decided to limit our investigation to the upper left limb as it is one of the most agile and frequently-used body parts and with the consideration of the human labor cost in data collection.
However, we expect our approach to be applicable to similar investigations for other limbs and body parts to construct the offset model. 
This has considerable potentials in future applications in maintaining whole-body immersion while enabling body-scale interaction functionalities and thus will be further explored in future work.

In Section~\ref{section:study1} and Section~\ref{section:study2}, we conducted four user studies with 12 participants for each, which results in 48 participants in total.
Considering the long experiment time (around 90 minutes), heavy physical load of task, it is hard to conduct studies with massive users, especially during the COVID-19 pandemic.
We conducted repeated measures ANOVA tests with the variables of user on the proportion of noticing the offset for the same set of poses and offsets and the results showed that user did not significantly affect the noticing probability $(p > 0.05)$ which indicated that these users' data were similar and thus the 12 users' data are appropriate to construct a statistical model.
We constructed the model based on 36 users' data that we collected in the three phases in Section~\ref{section:study1} and evaluated it with 12 users in the evaluation study.
We recognize that our model's prediction accuracy is constricted by the number of participants and the accuracy could be better with more participants.
The evaluation results were promising which indicate that the model constructed with relatively small data worked well.
We will repeat the studies with more users and validate our conclusions and model in the future.

In the evaluation in Section~\ref{section:study2}, we compared three offset distributions between the shoulder and the elbow joints to study how offset distribution affects interaction efficiency and experience in Section \ref{section:study2} and found that balanced offsets were optimal for immersion. 
We chose three representative distributions ($\delta_{s}$ - $\delta_{s}$: 1 - 0, 0.5 - 0.5, 0 - 1) and three strength level(none, medium, and high).
We only tested three conditions for each of the two factors cause we were not sure whether these two factors would significantly affect the performances. 
Therefore we conducted an elementary explorative experiment and we would test more levels, for instance, offset distribution of (0.75 - 0.25) or (0.25 - 0.75). 
We recognised that the evaluation of distributions can be extended to identify other optimal offset strength and distribution.

We investigated the movement inconsistency's noticeability from the first-person point of view in this paper.
However, existing research found that users could observe their motions with more clarity with a third-person point of view~\cite{juyong2020effects, albert2019the} in VR.
We expect the trend of body ownership degradation with increased offset differs from that of the first-person point of view and is worth investigating.

\section{Conclusion}
\label{section:conclusion}

This paper explores how the strength of user-avatar movement inconsistency affects its noticeability and finally the sense of body ownership in VR. 
In approaching this, we investigated the effect of applying angular offsets to the left arm's shoulder and elbow on body ownership by the probability of noticing the existence of the offset. 
We conducted three user studies to quantify the effect of offset on the offset noticing probability.
With this quantified effect, we can predict the probability of noticing the offset with a given composite two-joint offset at various pose. Leveraging this knowledge, we then constructed a statistical model which outputs a set of applicable composite two-joint offsets given an offset noticing probability and an arm pose. 
We adapted our model to construct dynamic movement with a maximum $75\%$ offset noticeability to support an interaction technique optimizing movement amplification. 
We evaluated the technique, and the results show that the offset could significantly reduce the moving time and distance while a medium-level offset does not compromise the sense of body ownership.
We demonstrated the extendability and usability of the proposed model in three applications with different body ownership requirements. 
Finally, we discussed other factors that affected body ownership as possible future work and recognized the limitations.

\section*{Acknowledgments}
This work is supported by the Natural Science Foundation of China under Grant No.6213000120, 62002198, Tsinghua University Initiative Scientific Research Program, the China Postdoctoral Science Foundation	under Grant No.2021M691788, Key Research Projects of the Foundation Strengthening Program under Grant No. 2020JCJQZD01412, the Natural Science Foundation of China under Grant No. 62132010,
Beijing Key Lab of Networked Multimedia, and the Institute for Guo Qiang and Institute for Artifcial Intelligence, Tsinghua University.

\bibliographystyle{ACM-Reference-Format}
\bibliography{main}


\end{document}